%% file: paper2.tex
\documentclass[12pt]{article}
\usepackage{amssymb}
\usepackage{amsopn}

\def\hybrid{\topmargin 0pt	\oddsidemargin 0pt 
	\headheight 0pt	\headsep 0pt
	\textheight 9in		
	\textwidth 6.1in	
	\marginparwidth .875in
	\parskip 5pt plus 1pt	\jot = 1.5ex}

\hybrid

\catcode`\@=11

\def\numberbysection{\@addtoreset{equation}{section}
	\def\theequation{\thesection.\arabic{equation}}}

\def\titlepage{\@restonecolfalse\if@twocolumn\@restonecoltrue\onecolumn
     \else \newpage \fi \thispagestyle{empty}\c@page\z@	
	\def\thefootnote{\fnsymbol{footnote}} }

\def\endtitlepage{\if@restonecol\twocolumn \else \newpage \fi
	\def\thefootnote{\arabic{footnote}} 
	\setcounter{footnote}{0}}  

\catcode`@=12
\relax

\numberbysection

\input{Ricci}

\DeclareMathOperator{\Orth}{O}
\DeclareMathOperator{\SOrth}{SO}

\DeclareMathOperator{\adtil}{\widetilde{ad}}
\DeclareMathOperator{\ad}{ad}

\DeclareMathOperator{\so}{\mathfrak{s}\mathfrak{o}}
\newcommand{\Btil}{\widetilde{B}}
\newcommand{\Ctil}{\widetilde{C}}
\newcommand{\Etil}{\widetilde{E}}
\newcommand{\Gtil}{\widetilde{G}}
\newcommand{\Htil}{\widetilde{H}}
\newcommand{\Mtil}{\widetilde{M}}

\newcommand{\Pitil}{\widetilde{\Pi}}

\newcommand{\Ttil}{\widetilde{T}}

\newcommand{\atil}{\tilde{a}}
\newcommand{\bbR}{\mathbb{R}}
\newcommand{\brtil}[2]{[#1,#2]^{\sim}}
\newcommand{\calB}{\mathcal{B}}
\newcommand{\coboundary}{\partial^{*}}
\newcommand{\ctil}{\tilde{c}}

\newcommand{\etil}{\tilde{e}}

\newcommand{\fhat}{\hat{f}}
\newcommand{\framebundle}{\mathcal{F}(P)}
\newcommand{\ftil}{\tilde{f}}

\newcommand{\gtil}{\tilde{g}}
\newcommand{\half}{\frac{1}{2}}

\newcommand{\lambdatil}{\tilde{\lambda}}
\newcommand{\liegtil}{\tilde{\mathfrak{g}}}
\newcommand{\lieg}{\mathfrak{g}}
\newcommand{\liek}{\mathfrak{k}}
\newcommand{\ltil}{\tilde{l}}

\newcommand{\myLie}[1]{\mathcal{L}_{#1}}
\newcommand{\ntil}{\tilde{n}}
\newcommand{\nutil}{\tilde{\nu}}

\newcommand{\psikperp}{\psi^{\liek^{\perp}}}
\newcommand{\psik}{\psi^{\liek}}

\newcommand{\qtil}{\tilde{q}}

\newcommand{\sigmatil}{\tilde{\sigma}}
\newcommand{\sleft}{[\![}
\newcommand{\sright}{]\!]}
\newcommand{\tautil}{\tilde{\tau}}
\newcommand{\thetatil}{\tilde{\theta}}
\newcommand{\tortil}{\tilde{\tau}^{\liek^{\perp}}}
\newcommand{\tor}{\tau^{\liek^{\perp}}}

\newcommand{\oaref}[2]{I-#2}

\begin{document}
\bibliographystyle{utphys}
\begin{titlepage}
\noindent
\strut\mbox{March 2000}\hfill{UMTG--222}\newline
\strut\hfill{\tt hep-th/0003178}
\par\vskip 2cm
\begin{center}
    {\Large \bf Target Space Duality II: Applications\footnote{This work
    was supported in part by National Science Foundation grant
    PHY--9870101.}}\\[0.5in]
    {\bf Orlando Alvarez}\footnote{email: \tt oalvarez@miami.edu}\\[0.1in]
    Department of Physics\\
    University of Miami\\
    P.O. Box 248046\\
    Coral Gables, FL 33124
\end{center}
\par\strut\vspace{.5in}
\noindent

\begin{abstract}
    We apply the framework developed in Target Space Duality~I: 
    General Theory. We show that both nonabelian duality and 
    Poisson-Lie duality are examples of the general theory. We propose 
    how the formalism leads to a systematic study of duality by 
    studying few scenarios that lead to open questions in the theory 
    of Lie algebras. We present evidence that there are probably
    new examples of irreducible target space duality.
\end{abstract}

\vspace{.5in}

\end{titlepage}

\section{Introduction}

In this article we consider some applications of the general theory
derived in article~I~\cite{Alvarez:2000bh}.  We show that nonabelian
duality 
\cite{Fridling:1984ha,Fradkin:1985ai,Kiritsis:1991zt,%
Rocek:1992ps,Giveon:1992jj,%
delaOssa:1993vc,Gasperini:1993nz,%
Giveon:1994mw,Giveon:1994ai,Giveon:1994ph} and Poisson-Lie duality
\cite{Klimcik:1995ux,Klimcik:1996dy,Klimcik:1996kw,Sfetsos:1998pi} are
special cases of the general theory.  In fact we show that nonabelian
duality is a special case of a more general situation.  The spirit of
this paper is that there are natural geometric scenarios that need to
be explored.  We explore a few of the easier ones and see that they
lead to open mathematical questions in the theory of Lie algebras. 
For example, Lie bialgebras and generalizations, and $R$-matrices
naturally appear in this framework.  References to equations and
sections in article~I are preceded by~I, \emph{e.g.}, (I-8.3).

\section{Examples with a flat $\psi$ connection}
\label{sec:examplespsi}

\subsection{General remarks}
\label{sec:psiremarks}
\relax
To best understand how to use the equations given in the
Section~\oaref{sec:psi}{8.1} it is best to do a few examples.  The examples
we will consider in Section~\ref{sec:examplespsi} assume that the
connection $\psi_{ij}$ is flat.  We are mostly interested in local
properties so we might as well assume $P$ is parallelizable.  We can
use parallel transport with respect to this connection to get a global
framing.  In this special framing the connection coefficients vanish
and thus we can make the substitution $\psi_{ij}=0$ in all the
equations in Section~\oaref{sec:psi}{8.1}.  A previous remark tells us that
in this case $f_{ijk}$ and $\ftil_{ijk}$ are pullbacks respectively of
tensors on $M$ and $\Mtil$.  Consequently we have that
$df_{ijk}=f'_{ijkl}\theta^{l}$ and
$d\ftil_{ijk}=\ftil''_{ijkl}\thetatil^{l}$, \emph{i.e.},
$f''_{ijkl}=\ftil'_{ijkl}=0$.  Because we are interested in mostly
local considerations we might as well assume both $M$ and $\Mtil$ are
parallelizable.

\subsection{The tensor $n_{ij}$ is the pullback of a tensor on $\widetilde{M}$}
\label{sec:pullback}

Here we assume that the connection $\psi$ is flat.
In Section~\oaref{sec:covconst}{8.2} we considered what happened if $n_{ij}$ 
was covariantly constant in this section we relax this condition to one where 
$n_{ij}$ only depends on a natural subset of the variables. We assume 
that $n_{ij}$ is the pullback of a tensor on $\Mtil$. This means that
\begin{equation}
    dn_{ij}=n''_{ijk}\thetatil^{k}\quad\mbox{or equivalently}\quad
    n'_{ijk}=0\;.
    \label{eq:npullback}
\end{equation}
An equivalent formulation is that $\half
n_{ij}\thetatil^{i}\wedge\thetatil^{j}$ is the pullback of a $2$-form
on $\Mtil$.  We immediately see from (\oaref{eq:fnpsi}{8.17}) and
(\oaref{eq:ftilnpsi}{8.18}) that
\begin{eqnarray}
    n''_{ijk} & = & m_{kl}f_{lij}\,,
    \label{eq:mf}  \\
   n''_{kij}-n''_{kji} & = & m_{jl}f_{lki}-m_{il}f_{lkj} 
   = \ftil_{lij}m_{lk}\;.
    \label{eq:mftil}
\end{eqnarray}
A brief computation shows that $d(m_{ji}\thetatil^{j})=0$ and
therefore we can locally find $n$ functions $p_{1},\ldots,p_{n}$ such
that 
\begin{equation}
    dp_{i}=m_{ji}\thetatil^{j}\,.
    \label{eq:dpmtheta}
\end{equation}
Note that since $m_{ij}$ is invertible the differentials
$\{dp_{1},\ldots,dp_{n}\}$ are linearly independent.  The functions
$p_{1},\ldots,p_{n}$ are the pullbacks of functions locally defined on
$\Mtil$ and they define a local coordinate system on $\Mtil$.  We
immediately see that
\begin{equation}
    dn_{ij}=f_{kij}dp_{k}\,.  
    \label{eq:dnfdp}
\end{equation}
The integrability
condition $0=d^{2}n_{ij}=f'_{kijl}\theta^{l}\wedge dp_{k}$ immediately
tells us that $f'_{kijl}=0$ since
$\{\theta_{1},\ldots\theta_{n},dp_{1},\ldots, dp_{n}\}$ are linearly
independent.  Since $f'_{ijkl}=f''_{ijkl}=0$ we see that $f_{ijk}$ are
constants and thus $M$ is diffeomorphic to a Lie group $G$.  The
$\theta^{i}$ are the pullbacks by $\Pi$ of the left invariant
Maurer-Cartan forms on $G$.  The next question is whether the pullback
metric $\theta^{i}\otimes\theta^{i}$ is invariant under the action of
the group.  Choose a $G$ left-invariant vector field $X$ ``along'' $M$,
\emph{i.e.}, $\iota_{X}\thetatil^{i}=0$.  A brief computation shows
that
\begin{equation}
    \myLie{X}(\theta^{i}\otimes\theta^{i}) = X^{k}f_{ijk}
    (\theta^{i}\otimes\theta^{j} + \theta^{j}\otimes\theta^{i})
    \label{eq:Liemetric}
\end{equation}
The metric on $M$ is $G$-invariant if and only if $f_{ijk}=-f_{jik}$,
\emph{i.e.}, the structure constants are totally antisymmetric, a 
standard result.

We can integrate (\ref{eq:dnfdp}) to obtain
\begin{equation}
    n_{ij}=n^{0}_{ij} + f_{kij}p_{k}
    \label{eq:nintegral}
\end{equation}
where $n^{0}_{ij}$ are constants.  Define the constant tensor $m^{0}$
by $m^{0}_{ij}=\delta_{ij}+ n^{0}_{ij}$.  Equation~(\ref{eq:mftil})
immediately gives us an expression for $\ftil_{ijk}$ in terms of
$f_{ijk}$ and $p_{l}$.  It is a straightforward computation to verify
that $\ftil_{ijk}$ satisfies the integrability conditions for
$d\thetatil^{l}=-\half \ftil_{lij}\thetatil^{i}\wedge\thetatil^{j}$. 
In general $\ftil_{ijk}$ are not constants and are thus not 
generically Maurer-Cartan forms for a Lie group.

To determine $H$ we use (\oaref{eq:Hnppsi}{8.15}). Note that our hypotheses 
imply that the left hand side is automatically zero. After using the 
Jacobi identity satisfied by the $f_{ijk}$ we find that
\begin{equation}
    H_{ijk}= f_{lij}n^{0}_{lk}+f_{ljk}n^{0}_{li}+f_{lki}n^{0}_{lj}\;.
    \label{eq:Hresult}
\end{equation}
If we define a left invariant $2$-form by $n^{0}=\half 
n^{0}_{ij}\theta^{i}\wedge\theta^{j}$ then the above equation is
$H=-dn^{0}$ which tells us that $H$ is cohomologically trivial. We 
use (\oaref{eq:Htilnpppsi}{8.16}) to determine $\Htil$:
\begin{eqnarray}
    \Htil_{ijk} & = & (\ftil_{ijk}+\ftil_{jki}+\ftil_{kij})
    \nonumber  \\
     & + & H_{ijk} - (f_{ijk}+f_{jki}+f_{kij})\;.
    \label{eq:Htilresult}
\end{eqnarray}

\subsubsection{Cotangent bundle duality}
\label{sec:cotangent}

We will see in this section that cotangent bundle duality is a special
case of what we have been discussing in Section~\ref{sec:pullback}. 
In particular this corresponds to what is often called ``nonabelian
duality''.  The cotangent bundle has a natural symplectic structure
and thus we automatically have a candidate symplectic manifold $P$ for
free.  Assume the manifold $M$ is parallelizable.  This means that the
cotangent bundle $T^{*}M$ is trivial, \emph{i.e.}, it is a product
space, $P=T^{*}M = M\times\bbR^{n}$.  The projections $\Pi$ and
$\Pitil$ will be taken to be the cartesian projections and therefore
$\Mtil=\bbR^{n}$.  If $\theta^{i}$ is an orthonormal coframe on $M$
then the canonical $1$-form on $T^{*}M$ may be written as
$\alpha=p_{i}\theta^{i}$.  The $p_{i}$ are coordinates along the
fibers of $\Pi$.  The fibers of $\Pitil$ (diffeomorphic to $M$) given
by $dp_{i}=0$, $i=1,\ldots,n$ are the same as the fibers given by
$\thetatil^{i}=0$, $i=1,\ldots,n$ therefore there must exist an
invertible matrix defined by functions $\widehat{m}_{ij}$ such that
\begin{equation}
    dp_{j}=\widehat{m}_{ij}\thetatil^{i}\;.
    \label{eq:dptothetatil}
\end{equation}
The canonical symplectic form is given by
$\beta=d\alpha=dp_{j}\wedge\theta^{j}-\half
p_{k}f_{kij}\theta^{i}\wedge\theta^{j}$.  In going from $\beta$ to
$\gamma$ the $\thetatil\wedge\theta$ term is not changed and so we
immediately learn that $\widehat{m}_{ij}=m_{ij}=\delta_{ij}+n_{ij}$. 
Taking the exterior derivative of (\ref{eq:dptothetatil}) leads to
$n'_{ijk}=0$ and (\ref{eq:mftil}).  Thus $n_{ij}$ is the pullback of a
tensor on $\bbR^{n}$ and we are back to our general discussion given
in Section~\ref{sec:pullback}.  This is an example of what is often
called ``nonabelian duality'' which has generating function given by
(\oaref{eq:defF}{2.2}) with $\alpha=p_{k}\theta^{k}$.  We know that $M$ is a
Lie group $G$, $T^{*}G = G\times \lieg^{*}$ and thus $\Mtil=
\lieg^{*}$.  The metric on $\lieg^{*}$ is immediately computable from
(\ref{eq:dptothetatil}) since
$m_{ij}=\delta_{ij}+n^{0}_{ij}+p_{k}f_{kij}$.  It is worth remarking
that because $\Mtil=\lieg^{*}=\bbR^{n}$ there is an abelian Lie group
action on $\Mtil$ which does not leave the metric invariant.  Said
differently the $dp$ are the Maurer-Cartan forms for the abelian Lie
group $\lieg^{*}$.  This statement is made in anticipation of our
discussion about Poisson-Lie duality in Section~\ref{sec:Poisson-Lie}.

``Nonabelian duality'' usually refers to the case with $n^{0}_{ij}=0$
where (\ref{eq:Hresult}) tells us that $H_{ijk}=0$.  We can compute
$\Htil$ from (\ref{eq:Htilresult}) directly or it was already remarked
in Section~\oaref{sec:Liegroup}{7.2} that $\Btil_{ij}$ is easily determined.

\subsubsection{Cotangent bundle duality and gauge invariance}
\label{sec:noncartesian}

In this section we revisit cotangent bundle duality and try to
understand geometrically the role of the $B$ field gauge
transformation.  We assume the manifold $M$ has a trivial cotangent
bundle $T^{*}M= M\times \bbR^{n}$ with canonical $1$-form $\alpha =
p_{i}\theta^{i}$ and symplectic form $\beta=d\alpha$.  We modify the
discussion of Section~\ref{sec:cotangent} by demanding that the
bifibration not be given by the cartesian projections.  We want the
vertical fibers to be original ones so we have $\Pi:(x,p)\in T^{*}M
\mapsto x\in M$.  On the other hand the projection
$\Pitil:T^{*}M\to\Mtil$ will not be the canonical projection.  The
fibers of this projection are ``slanted'' relative to the fibers of
the cartesian projection $\Pitil_{c}$.  Note that the fibers of
$\Pitil$ are the integral manifolds of the Pfaffian equations
$\thetatil^{i}=0$.  From general principles we know that
\begin{equation}
    dp_{i} = \widehat{m}_{ji}\thetatil^{j} + u_{ji}\theta^{j}
    \label{eq:dpnc}
\end{equation}
for some functions $\widehat{m}, u$.  Geometrically this is the
statement that the fibers of $\Pitil$ are slanted relative to the
fibers of $\Pitil_{c}$.  As before the structure of the symplectic
form leads to the result that $\widehat{m}_{ij}=m_{ij}$.  The
integrability condition $d^{2}p_{i}=0$ leads to the following
equations
\begin{eqnarray}
    n''_{ijk}-n''_{ikj} & = & \ftil_{ljk}m_{li}\,,
    \label{eq:cot1}  \\
    u'_{jik}-u'_{kij} & = & -f_{ljk}u_{li}\,,
    \label{eq:cot2}  \\
    u''_{jik} & = & n'_{kij}\,.
    \label{eq:cot3}
\end{eqnarray}
Comparing (\ref{eq:cot1}) with (\oaref{eq:ftilnpsi}{8.18}) we see that
$n'_{ijk}=0$ and thus $n_{ij}$ is the pullback of a tensor on $\Mtil$. 
The general discussion of Section~\ref{sec:pullback} tells us that
there exists function $\hat{p}_{i}$ in $T^{*}M$ such that
$d\hat{p}_{i}=m_{ji}\thetatil^{j}$.  Geometrically this is just the
statement that the fibers is given by
$\hat{p}_{i}=\mbox{constant}$.  Note that (\ref{eq:cot3}) tells us
that $u''_{ijk}=0$ and thus $u_{ij}$ is the pullback of a tensor on
$M$.  Equation~(\ref{eq:dpnc}) tells us that $d(p_{i}-\hat{p}_{i}) =
u_{ji}\theta^{j}$ and thus we see that $p_{i}=\hat{p}_{i} + k_{i}$
where $k_{i}$ are pullbacks of functions on $M$, \emph{i.e.},
$k_{i}=k_{i}(x)$. Thus we we see that the canonical transformation 
generated by $\int p_{i}\theta^{i}$ is gauge 
equivalent to the one generated by $\int \hat{p}_{i}\theta^{i}$ and 
corresponds to a different choice of fibration.

\subsubsection{Can $\widetilde{M}$ naturally be a Lie group?}
\label{sec:dualLiegroup}

Is it possible for the dual manifold to be \emph{naturally} a Lie
group under the assumptions underlying the discussions in
Section~\ref{sec:pullback}?  By ``naturally'' we mean that the
$\thetatil^{i}$ are the Maurer-Cartan forms for a Lie group $\Gtil$. 
Solving (\ref{eq:mftil}) for $\ftil_{ijk}$ generally leads to a
non-constant solution.  There is a possibility that the solution may
be constant, \emph{i.e.}, $(\Mtil,\gtil,\Btil)$ is naturally a Lie
group.  Unfortunately we will see that both $\lieg$ and $\liegtil$
must be abelian and so there are no new interesting examples. 
Inserting (\ref{eq:nintegral}) into (\ref{eq:mftil}) and using the
linear independence of the $dp_{i}$ leads to two equations:
\begin{eqnarray}
    m^{0}_{jl}f^{l}{}_{ki}-m^{0}_{il}f^{l}{}_{kj} 
    & = & \ftil^{l}{}_{ij}m^{0}_{lk}\;,
    \label{eq:lie1}  \\
    f^{m}{}_{il}f^{l}{}_{jk} & = & f^{m}{}_{il}\ftil^{l}{}_{jk}\;.
    \label{eq:lie2}
\end{eqnarray}
To obtain the latter equation we used the Jacobi identity satisfied by
$f_{ijk}$.  The indices have also been set at their natural
(co)variances for future convenience.  The $\ftil^{i}{}_{jk}$ satisfy 
the Jacobi identity if they satisfy the two equations above because of
our remark about integrability after (\ref{eq:nintegral}).
You should think of $M$ as a
Lie group $G$ with Lie algebra $\lieg$ and similarly for $\Mtil$.  Let
$\ad_{X}(Y)=[X,Y]$ be the adjoint action of $\lieg$.  Let
$\brtil{\cdot}{\cdot}$ be the Lie bracket on $\liegtil$ with adjoint
action denoted by $\adtil_{\widetilde{X}}{\widetilde{Y}} =
\brtil{\widetilde{X}}{\widetilde{Y}}$.  The reduction of the structure
group of $P$ to $\Orth(n)$ meant that we could identify the horizontal
tangent space with the vertical tangent space and thus we can identify
$\lieg$ with $\liegtil$.  We should think of a single vector space $V$
with two Lie brackets giving us two Lie algebras
$\lieg=(V,[\cdot,\cdot])$ and $\liegtil=(V,\brtil{\cdot}{\cdot})$.  In
this notation (\ref{eq:lie2}) becomes
\begin{equation}
    \ad_{X}(\ad_{Y} -\adtil_{Y}) = 0\,.
    \label{eq:lie3}
\end{equation}
There are some immediate important consequences of this equation.  Let
$\mathfrak{d}$ be the vector subspace of $\lieg$ spanned by $\ad_{Y}Z
- \adtil_{Y}Z$ for all $Y,Z$.  The subspace $\mathfrak{d}$ is
contained in the center $\mathfrak{z}$ of $\lieg$ by (\ref{eq:lie3}). 
If $\mathfrak{d} \neq 0$, \emph{i.e.}, $\ad\neq\adtil$, then the
center $\mathfrak{z}$ is a nontrivial abelian ideal in the Lie
algebra $\lieg$ and thus $\lieg$ is not semisimple.  If
$\mathfrak{d}=0$ then we have that $\ad=\adtil$ and we will show that
$\lieg$ is abelian when we incorporate (\ref{eq:lie1}) into our
reasoning.  Note that if $\mathfrak{z}=0$ then $\mathfrak{d}=0$.

Let us study the implications of $\ad=\adtil$, \emph{i.e.},
$\mathfrak{d}=0$.  Equation~(\ref{eq:lie1}) may be rewritten as
\begin{equation}
    2f_{kij}=(f_{ijk}+f_{jki}+f_{kij}) + 
    (n^{0}_{il}f_{ljk}+n^{0}_{jl}f_{lki}+n^{0}_{kl}f_{lij})\;.
    \label{eq:fskew}
\end{equation}
Notice that the right hand side is totally antisymmetric under
permutations of $i,j,k$.  This means that $f_{kij}$ is totally
antisymmetric and thus the metric $\theta^{i}\otimes\theta^{i}$ is a
bi-invariant positive definite metric on $G$.  The proposition proven
in Appendix~\ref{sec:Liegroupprimer} tells us that there is a
decomposition into ideals $\lieg=\liek \oplus \mathfrak{a}$ where
$\liek$ is a compact semisimple Lie algebra and $\mathfrak{a}$ is an
abelian Lie algebra.  The next part of the argument only involves the
compact ideal $\liek$ and without any loss we assume $\mathfrak{a}=0$
for the moment.  Using the antisymmetry of the structure constants the
above may be rewritten as
$$
    -f_{ijk}=
    n^{0}_{il}f_{ljk}+n^{0}_{jl}f_{lki}+n^{0}_{kl}f_{lij}\;.
$$
Using the remark made immediately after (\ref{eq:Hresult}) we see that
$\frac{1}{3!}f_{ijk}\theta^{i}\wedge\theta^{j}\wedge\theta^{k} =
-dn^{0}$.  The closed $3$-form
$\frac{1}{3!}f_{ijk}\theta^{i}\wedge\theta^{j}\wedge\theta^{k}$ is
exact and this contradicts $H^{3}(\liek)\neq 0$ as discussed in
Appendix~\ref{sec:Liegroupprimer}.  If $\ad=\adtil$ then we conclude
that $\lieg=\mathfrak{a}$ is abelian and we are back in familiar territory.

The next we consider the general case by exploiting the observation
that $\mathfrak{d}\subset\mathfrak{z}$.  Since we have a positive
definite metric on $\lieg$ there is an orthogonal direct sum
decomposition $\lieg = \mathfrak{z} \oplus \mathfrak{z}^{\perp}$.  The
orthogonal complement $\mathfrak{z}^{\perp}$ is not generally an ideal
because the metric is not necessarily $(\ad\lieg)$-invariant. 
Nevertheless we can choose an orthonormal basis $\{e_{\alpha}\}$ for
$\mathfrak{z}$ and an orthonormal basis $\{e_{a}\}$ for
$\mathfrak{z}^{\perp}$.  Greek indices are associated with
$\mathfrak{z}$, indices in $\{a,b,c,d\}$ are associated with
$\mathfrak{z}^{\perp}$; and indices in $\{i,j,k,l\}$ run from
$1,\ldots,\dim\lieg$ and are associated with all of $\lieg$.  The Lie
algebra $\lieg$ is given by
\begin{eqnarray}
    [e_{\alpha},e_{j}] & = & 0\,,
    \label{eq:liegz}  \\
    \relax
    [e_{a},e_{b}] & = & f^{c}{}_{ab}e_{c} + 
    f^{\gamma}{}_{ab}e_{\gamma}\,.
    \label{eq:liegzperp}
\end{eqnarray}
Note that $\mathfrak{h} = \lieg/\mathfrak{z}$ is a Lie algebra since
$\mathfrak{z}$ is an ideal.  It follows that the structure constants
of $\mathfrak{h}$ are $f^{c}{}_{ab}$.  Also, $\lieg$ is a central
extension of $\mathfrak{h}$ with extension cocycle
$f^{\gamma}{}_{ab}$.  It is well known that the cocycle is trivial if
$f^{\gamma}{}_{ab} = t^{\gamma}{}_{c} f^{c}{}_{ab}$ corresponding to a
redefinition of the basis given by $e_{a} \to e_{a} +
t^{\gamma}{}_{a}e_{\gamma}$.  The Lie algebra $\liegtil$ must have the
form below because the two Lie brackets are the same modulo the center
$\mathfrak{z}$:
\begin{eqnarray}
    \brtil{e_{\alpha}}{e_{j}} & = & \ftil^{\gamma}{}_{\alpha j} e_{\gamma}\,,
    \label{eq:liegtilz}  \\
    \relax
    \brtil{e_{a}}{e_{b}} & = & f^{c}{}_{ab}e_{c} + 
    \ftil^{\gamma}{}_{ab}e_{\gamma}\,.
    \label{eq:liegtilzperp}
\end{eqnarray}
In equation~(\ref{eq:lie1}) choose $k=\gamma$ then the left hand side 
vanishes and the equation becomes $0=\ftil^{l}{}_{ij}m^{0}_{l\gamma} = 
\ftil^{d}{}_{ij}m^{0}_{d\gamma}+\ftil^{\delta}{}_{ij}m^{0}_{\delta\gamma}$. 
Using the different choices for $(i,j)$ we find
\begin{eqnarray}
    \ftil^{\delta}{}_{\alpha j} & = & 
    -\ftil^{d}{}_{\alpha j}m^{0}_{d\gamma}((m^{0})^{-1})^{\gamma\delta}=0\,,
      \nonumber \\
     \ftil^{\delta}{}_{ab} & = & 
    -f^{d}{}_{ab}m^{0}_{d\gamma}((m^{0})^{-1})^{\gamma\delta}\,.
    \label{eq:fcocycle}
\end{eqnarray}
These equations tell us that $\liegtil$ is a central extension of 
$\mathfrak{h}$ with a trivial cocycle. Next choose $i=\alpha$ in 
(\ref{eq:lie1}) with result $0 = m^{0}_{\alpha l}f^{l}{}_{kj}$. Choosing 
$k=a$ and $j=b$ leads to 
\begin{equation}
    f^{\gamma}{}_{ab} =
    -((m^{0})^{-1})^{\gamma\alpha} m^{0}_{\alpha c}f^{c}{}_{ab}=0
    \label{eq:ftilcocycle}
\end{equation}
which tells us that the cocycle $f^{\gamma}{}_{ab}$ is also trivial. 
Finally choose $(i,j,k) = (a,b,c)$ in (\ref{eq:lie1}) and substitute
(\ref{eq:fcocycle}) and (\ref{eq:ftilcocycle}) for
$\ftil^{\delta}{}_{ab}$ and $f^{\gamma}{}_{ab}$ to obtain
\begin{eqnarray}
    \left(m^{0}_{bd}-m^{0}_{b\gamma}((m^{0})^{-1})^{\gamma\alpha}m^{0}_{\alpha
    d}\right) f^{d}{}_{ca} & - &
    \left(m^{0}_{ad}-m^{0}_{a\gamma}((m^{0})^{-1})^{\gamma\alpha}m^{0}_{\alpha
    d}\right) f^{d}{}_{cb} \nonumber \\
    & = &
    f^{d}{}_{ab}
    \left(m^{0}_{dc} -  
    m^{0}_{d\alpha}((m^{0})^{-1})^{\alpha\gamma}m^{0}_{\gamma c}\right)\;.
    \label{eq:msoln}
\end{eqnarray}
Next we observe that since
$m^{0}_{\alpha\beta}=\delta_{\alpha\beta}+n_{\alpha\beta}$ we conclude
that $((m^{0})^{-1})^{\alpha\beta}= s^{\alpha\beta} + a^{\alpha\beta}$
where $s^{\alpha\beta}$ is symmetric and positive definite and
$a^{\alpha\beta}$ is skew.  In particular note that $ -m^{0}_{b
\gamma}((m^{0})^{-1})^{\gamma\alpha}m^{0}_{\alpha d}=
s^{\gamma\alpha}m^{0}_{\gamma b}m^{0}_{\alpha d} +
a^{\gamma\alpha}m^{0}_{\gamma b}m^{0}_{\alpha d}$ where the first term
is symmetric and positive definite and the second is skew.  Using this
we immediately see that
$$
    m^{0}_{bd}-m^{0}_{b\gamma}((m^{0})^{-1})^{\gamma\alpha}m^{0}_{\alpha d} =
    S_{bd}+N_{bd}
$$
where $S_{bd}$ is symmetric and positive definite and
$N_{bd}$ is skew.  We can use $S_{ab}$ as a second metric on $\lieg$
and use it to ``raise and lower'' the indices in (\ref{eq:msoln})
leading to
\begin{equation}
    2f_{cab} = (f_{abc}+f_{bca}+f_{cab}) + 
    \left( N_{ad}f^{d}{}_{bc}+N_{bd}f^{d}{}_{ca}+N_{cd}f^{d}{}_{ab} \right).
    \label{eq:fabc}
\end{equation}
This tells us that $f_{abc}$ is totally antisymmetric and therefore
$S_{ab}$ is an invariant metric on $\mathfrak{h}$.  The same chain of
arguments used after (\ref{eq:fskew}) tell us that $\mathfrak{h}$ is
abelian which implies that $\ftil^{\delta}{}_{ab}=0$ and
$f^{\gamma}{}_{ab}=0$ concluding that $\lieg$ and $\liegtil$ are
abelian Lie algebras.

\subsubsection{Connection with $R$-matrices}
\label{sec:Rmatrixconnection}

Equation (\ref{eq:lie1}) is closely related to the theory of
$R$-matrices developed by Semenov-Tian-Shansky~\cite{Semenov:83aa}
which is different than the one developed by
Drinfeld~\cite{Drinfeld:83aa}.  The discussion here suggests that
Poisson-Lie groups may play a role in duality, see
Section~\ref{sec:Poisson-Lie}.  We observe that (\ref{eq:lie1}) may be
rewritten as
\begin{equation}
    (m^{0})^{-1}_{jl}f_{kil}-(m^{0})^{-1}_{kl}f_{jil} = 
    (m^{0})^{-1}_{jl}(m^{0})^{-1}_{km} 
    \ftil_{nlm}m^{0}_{ni}\;.
    \label{eq:MYB}
\end{equation}
Next we show that this equation describes a potential double Lie
algebra structure on $\lieg$ \emph{\`{a} la} Semenov-Tian-Shansky
(\ref{eq:deffR}).  Assume the Lie algebra $\lieg$ admits an invariant
metric $K$.  For example if the Lie algebra is semisimple then $K$ may
be taken to be the Killing metric.  Let us use the indices $a,b,c,d,e$
to denote generic components in a generic basis.  In terms of a basis
$(e_{1},\ldots,e_{n})$ the structure constants are given by
$[e_{a},e_{b}]=f^{c}{}_{ab}e_{c}$.  For the moment it is best to
forget about the orthonormal basis we were previously using before
because the metric $K$ may not be related to the previous metric.  The
components of the invariant metric are given by
$K_{ab}=K(e_{a},e_{b})$.  We will use $K$ to identify $\lieg$ with
$\lieg^{*}$, \emph{i.e.}, raise and lower indices.  The tensor with
components $K^{ab}$ is the inverse of the invariant metric,
\emph{i.e.}, the induced metric on $\lieg^{*}$.  The statement that
$K$ is $\lieg$-invariant is equivalent to $f_{abc}$ being totally
antisymmetric.  With these assumptions equation (\ref{eq:Rf}) may be
rewritten as
\begin{equation}
    (R^{d}{}_{e}K^{ea})f^{b}{}_{cd} - (R^{d}{}_{e}K^{eb})f^{a}{}_{cd}
    =K^{aa'}K^{bb'}K_{cc'}(f^{c'}{}_{a'b'})_{R}.
    \label{eq:deffbar}
\end{equation}
The indices of $R$ and $f$ have their natural (co)variances.

We are now going to compare (\ref{eq:deffbar}) with (\ref{eq:MYB}).  
Remember that (\ref{eq:MYB}) is valid in an orthonormal basis with 
respect to a specific metric on $\lieg$ which may not be related to 
the invariant metric $K$.  What we will have to do is express 
(\ref{eq:deffbar}) in the orthonormal basis but this will be simple 
because we adjusted all the (co)variances correctly in the equation 
above.  The indices $i,j,k,l,m,n$ refer to our orthonormal basis.  We 
raise and lower indices using the Kronecker delta tensor.  The only 
potential confusion is that we have to be careful and remember that 
$K_{ab}$ becomes $K_{ij}$ and the inverse invariant metric $K^{ab}$ 
becomes $K^{-1}_{ij}$.  The correspondence is made by choosing $R$ (in 
our orthonormal basis) to be defined by
\begin{equation}
    (m^{0})^{-1}_{ij}= R_{jl}K^{-1}_{li}\;.
    \label{eq:defR}
\end{equation}
The $R$-matrices we are considering are invertible because both 
$m^{0}$ and $K$ are always invertible. This is different that in the 
Drinfeld case where the $R$-matrix is skew adjoint and may not be 
invertible.
If you think of $K$ as a map $K:\lieg\to\lieg^{*}$ then the equation 
above is $(m^{0})^{t}=KR^{-1}:\lieg\to\lieg^{*}$ which suggests 
that $m^{0}$ should be interpreted as a map 
$m^{0}:\lieg^{*}\to\lieg$. Comparing the right hand sides of 
(\ref{eq:MYB}) and (\ref{eq:deffbar}) we see that
\begin{equation}
    \ftil_{ilm}R_{lj}R_{mk}=R_{il}(f_{ljk})_{R}\;.
    \label{eq:fRrel}
\end{equation}

If $R_{ij}$ satisfies the modified Yang Baxter equation 
(\ref{eq:STSMYB}) then $(f^{i}{}_{jk})_{R}$ are the structure 
constants of a Lie algebra $\lieg_{R}$ associated with a Lie group 
$G_{R}$. Let $(\mu^{1},\ldots,\mu^{n})$ be the left invariant 
Maurer-Cartan forms for $G_{R}$:
$$
    d\mu_{l}= -\half (f_{ljk})_{R}\,\mu_{j}\wedge\mu_{k}\;.
$$
Going to a different basis given by $\lambda_{i}=R_{il}\mu_{l}$ we 
see that
$$
   d\lambda_{i}=-\half \ftil_{ilm} \lambda_{l}\wedge\lambda_{m}\;.
$$
We conclude that if $R$ is a solution of the modified classical Yang 
Baxter equation then we can construct the Lie algebra $\tilde{\lieg}$ 
with structure constant $\ftil_{ijk}$ and an associated Lie group 
$\Gtil$.

In our situation we also have to impose a second
equation~(\ref{eq:lie2}) and this is very restrictive.  Our in-depth
analysis that eliminated all except the abelian case.  Logically,
there is the possibility of non-trivial solutions by the use of
$R$-matrices.  The reason is that using (\ref{eq:fRrel}) and
(\ref{eq:deffR}) we can rewrite (\ref{eq:lie2}) as
$$
    0 = f^{m}{}_{il}( f^{l}{}_{jk}R^{j}{}_{p}R^{k}{}_{q}
    -f^{r}{}_{nq}R^{n}{}_{p}R^{l}{}_{r}
    -f^{r}{}_{pn}R^{n}{}_{q}R^{l}{}_{r})\;.
$$
The stuff between the parentheses is the left hand side of modified
classical Yang Baxter equation (\ref{eq:STSYBcomp}).  If $R$ satisfies
the modified classical Yang Baxter equation the above becomes
$0=-cf^{m}{}_{il}f^{l}{}_{pq}$.  The solution to this equation is
$c=0$ or $\ad_{X}\ad_{Y}=0$. We know from our general analysis that it 
is unnecessary to proceed along these lines.

\subsection{Symmetric duality}
\label{sec:symmetricduality}

In the previous sections we had $\alpha = p_{i}\theta^{i}$ for some 
functions $p_{i}$ on $P$. A consequence was that the $p_{i}$ were good 
coordinates on the fibers of $\Pi$ and these fibers were also 
lagrangian submanifolds of $\beta$. There was a certain asymmetry in 
the way fibers of $\Pi$ and of $\Pitil$ were treated. In this section 
we consider a more symmetric situation. To motivate the ensuing 
presentation 
let us go to the case of $M=\bbR^{n}$ and $P=T^{*}M$ where we have the 
symplectic form $\beta = dp_{i}\wedge dq^{i}$. Up to constant 
$1$-forms the most general antiderivative is $\alpha = (1-u) 
p_{i}dq^{i} -u q^{i}dp_{i}$ where $u\in\bbR$ is a constant.

Let $U \subset P$ be a neighborhood, we can always write
\begin{equation}
    \alpha = \qtil_{i}\theta^{i} - q^{i}\thetatil_{i}\;,
    \label{eq:defqqtil}
\end{equation}
where $q^{i}$ and $\qtil_{i}$ are local functions on $U$.  We now make
some special assumptions about the functions $q$ and $\qtil$.  Assume
there exist matrix valued functions $E$ and $\Etil$ such that
\begin{eqnarray}
    dq^{i} & = & E_{ij}\theta^{j}\;,
    \label{eq:defE}  \\
    d\qtil_{j} & = & \thetatil_{i}\Etil_{ij} \;.
    \label{eq:defEtil}
\end{eqnarray}
These functions are not arbitrary and must satisfy a variety of
constraints.  For example, the ranks are constrained by equation
(\ref{eq:msym}) below.  For example if $E$ and $\Etil$ are invertible
then the functions $(q,\qtil)$ are independent in the sense that the
map $(q^{1},\ldots,q^{n},\qtil_{1}, \ldots,\qtil_{n}):U\to\bbR^{2n}$
is of rank $2n$.  Consequently the fibers of $\Pi$ are locally
described by $(q=\mbox{constant})$ and the fibers of $\Pitil$ by
$(\qtil=\mbox{constant})$.  In general $E$ or $\Etil$ will not be
invertible.  In this case not all the functions will be independent
and $(q=\mbox{constant})$ will define a family of manifolds each
diffeomorphic to $\Mtil$.

As before we use the prime and double prime notation to denote
derivatives in the appropriate directions.  The equation
$d^{2}q^{i}=0$ tells us that
\begin{eqnarray}
    E''_{ijk} & = & 0\;,
    \label{eq:Epp0}  \\
    E'_{ijk}-E'_{ikj} & = & -E_{il}f_{ljk}\;.
    \label{eq:Epf}
\end{eqnarray}
Likewise the equation $d^{2}\qtil=0$ tells us that 
\begin{eqnarray}
    \Etil'_{ijk} & = & 0\;,
    \label{eq:Etilp0}  \\
    \Etil''_{kjl}-\Etil''_{ljk} & = & -\ftil_{ikl}\Etil_{ij}\;.
    \label{eq:Etilppftil}
\end{eqnarray}
Note that you can solve (\ref{eq:Epf}) for $E'_{ijk}$ as a function 
of $E_{il}f_{ljk}$ and you can solve (\ref{eq:Etilppftil}) for 
$\Etil''_{ijk}$ and as function of $\ftil_{ikl}\Etil_{ij}$.

Next we compute $\beta=d\alpha$:
\begin{equation}
    \beta = (\Etil_{ij}+E_{ij})\thetatil^{i}\wedge\theta^{j}
    -\half\qtil_{i}f_{ijk}\theta^{j}\wedge\theta^{k}
    +\half  q^{i}\ftil_{ijk}\thetatil^{j}\wedge\thetatil^{k}\;.
    \label{eq:betasym}
\end{equation}
In general neither the fibers of $\Pi$ or of $\Pitil$ are lagrangian 
submanifolds of $\beta$. This is very different from the cotangent 
bundle cases previously discussed. 
Next we make we write $E$ and $\Etil$ as
\begin{eqnarray}
    E_{ij} & = & \sigma_{ij}+\nu_{ij}\;,
    \label{eq:defnu}  \\
    \Etil_{ij} & = & \sigmatil_{ij}+\nutil_{ij}\;
    \label{eq:defnutil}
\end{eqnarray}
where $\sigma$ and $\sigmatil$ are symmetric, and $\nu$ and $\nutil$
are antisymmetric.  Comparing to (\oaref{eq:gammareduced}{6.5}) 
we see that
\begin{equation}
    \delta_{ij}=\sigma_{ij}+\sigmatil_{ij}\quad\mbox{and}\quad
    n_{ij}=\nu_{ij}+ \nutil_{ij}\;.
    \label{eq:msym}
\end{equation}
The matrix valued functions $E$ and $\Etil$ cannot be arbitrary and 
must satisfy conditions given by the above.
We remark that 
$n'_{ijk}=m'_{ijk} =E'_{ijk}+\Etil'_{ijk} 
=E'_{ijk}= \sigma'_{ijk}+\nu'_{ijk}$. Since $n_{ij}=-n_{ji}$ we 
immediately see that
\begin{equation}
    \sigma'_{ijk} = 0\,, \quad\mbox{and}\quad 
    E'_{ijk}= n'_{ijk} =  \nu'_{ijk}\;.
    \label{eq:npnup}  
\end{equation}
Also $E''_{ijk}=0$ and thus we see that $\sigma''_{ijk}=0$ and 
$\nu''_{ijk}=0$. This means that $\sigma_{ij}$ is constant.
Similarly we conclude that $\sigmatil_{ij}$ is constant, 
$\Etil''_{ijk} =n''_{ijk} =\nutil''_{ijk}$ and $\nutil'_{ijk}=0$.

We can take the results given above and insert into 
(\oaref{eq:ftilnpsi}{8.18}) and (\oaref{eq:fnpsi}{8.17}) to obtain
\begin{eqnarray}
    E'_{ijk} =n'_{ijk}=\nu'_{ijk} &=& -\ftil_{lij}E_{lk}\;.
    \label{eq:unp} \\
    \Etil''_{ijk} = n''_{ijk}=\nutil''_{ijk} &=&  +\Etil_{kl}f_{lij}\;.
    \label{eq:1munpp}
\end{eqnarray}
We now insert the above into (\ref{eq:Epf}) and 
(\ref{eq:Etilppftil}) to obtain the basic equations
\begin{eqnarray}
    \Etil_{kl}f_{lij} - \Etil_{jl}f_{lik} & = & 
    \ftil_{ljk} \Etil_{li}\,,
    \label{eq:Etilf}  \\
    \ftil_{lij}E_{lk} - \ftil_{lik}E_{lj} & = & 
    E_{il} f_{ljk}\,.
    \label{eq:Eftil}
\end{eqnarray}
Note that 
\begin{eqnarray}
    dE_{ij}&=& E'_{ijk}\theta^{k}=\ftil_{lij}E_{lk}\theta^{k}= 
    \ftil_{lij}dq^{l}
    \label{eq:dE} \\
    d\Etil_{ij} &=& \Etil''_{ijk}\thetatil^{k} = 
    \Etil_{kl}f_{lij}\thetatil^{k} = f_{lij}d\qtil^{l}
    \label{eq:dEtil}
\end{eqnarray}
where we used (\ref{eq:defE}) and (\ref{eq:defEtil}).
We know that $df_{lij}= f'_{lijk}\theta^{k}$ and $d\ftil_{lij}= 
\ftil''_{lijk}\thetatil^{k}$. Therefore by taking the exterior 
derivatives of (\ref{eq:dE}) and (\ref{eq:dEtil}) we learn that
\begin{eqnarray}
    \Etil_{ml} (df_{lij}) & = & 0\,,
    \label{eq:df}  \\
    (d\ftil_{lij}) E_{lm} & = & 0\,.
    \label{eq:dftil}
\end{eqnarray}

To make progress we have to make some assumptions.  The simplest
assumption is that $E_{ij}=0$.  In this case we are back to the
discussion given in Section~\ref{sec:cotangent}.  In this paragraph we
use the notation that a Lie group $G$ via nonabelian duality is dual
to $\Gtil \approx \lieg^{*}$.  You can generalize cotangent bundle
duality along the following lines.  Consider matrices with with the
same $2\times 2$ block form
$$
    E = \left(
    \begin{array}{cc}
        E^{(1)} & 0  \\
        0 & 0
    \end{array}
    \right) , \qquad
    \Etil = \left(
    \begin{array}{cc}
        0 & 0  \\
        0 & \Etil^{(2)}
    \end{array}
    \right) .
$$
This will lead to a manifold $M = \Gtil^{(1)}\times G^{(2)}$ and 
$\Mtil = G^{(1)}\times \Gtil^{(2)}$.

Next we look at the case where $\Etil_{ij}$ is invertible. 
Equation~(\ref{eq:df}) tells us that $f_{ijk}$ are constants and thus
$M$ is naturally a Lie group $G$ with structure constants $f_{ijk}$. 
We can immediately integrate (\ref{eq:dEtil}) obtaining
\begin{equation}
    \Etil_{ij} = \sigmatil_{ij} + \nutil^{0}_{ij} + 
    f_{kij}\qtil^{k}\,,
    \label{eq:Etilsol}
\end{equation}
where $\nutil^{0}$ is a constant tensor.  With this information we can
use (\ref{eq:Etilf}) to determine $\ftil_{ijk}$.  It is
straightforward to verify that $\ftil_{ijk}$ satisfy the integrability
conditions for (\oaref{eq:cartan1psitil}{8.2}) with $\psi_{ij}=0$.  All we
have to do is to find a tensor $E_{ij}$ that satisfies
(\ref{eq:Eftil}) and (\ref{eq:dE}).  
Note that (\ref{eq:msym}) tells
us that $\sigma_{ij}=\delta_{ij} -\sigmatil_{ij}$.  This case merits 
further analysis.

If besides $\Etil_{ij}$ being invertible we also impose that
$E_{ij}$ is invertible then $\ftil_{ijk}$
are constant (see (\ref{eq:df})) and $\Mtil$ is naturally a Lie group
$\Gtil$.  We can integrate (\ref{eq:dE}) to obtain
\begin{equation}
    E_{ij} = \sigma_{ij} + \nu^{0}_{ij} + \ftil_{lij} q^{l}\,,
    \label{eq:Esol}
\end{equation}
where $\nu^{0}$ is a constant tensor. It is convenient to define
\begin{eqnarray}
    E^{0} & = & \sigma + \nu^{0},
    \label{eq:defE0}  \\
    \Etil^{0} & = & \sigmatil + \nutil^{0}.
    \label{eq:defEtil0}
\end{eqnarray}
We can insert (\ref{eq:Etilsol}) and (\ref{eq:Esol}) into 
(\ref{eq:Etilf}) and (\ref{eq:Eftil}) and expand both sides in powers 
of $q$ and $\qtil$ to obtain:
\begin{eqnarray}
    \Etil^{0}_{jl}f^{l}{}_{ki} - \Etil^{0}_{il}f^{l}{}_{kj} & = & 
    \ftil^{l}{}_{ij}\Etil^{0}_{lk}\,,
    \label{eq:ftoftil}  \\
    f^{m}{}_{il}f^{l}{}_{jk} 
    & = & f^{m}{}_{il}\ftil^{l}{}_{jk}\,,
    \label{eq:fftofftil}  \\
    \ftil^{l}{}_{ki}E^{0}_{lj} - \ftil^{l}{}_{kj}E^{0}_{li} & = & 
    E^{0}_{kl} f^{l}{}_{ij}\,,
    \label{eq:ftiltof}  \\
    \ftil^{m}{}_{il}\ftil^{l}{}_{jk}  & = & 
    \ftil^{m}{}_{il}f^{l}{}_{jk}\,,
    \label{eq:ftilftiltoftilf}
\end{eqnarray}
where the Jacobi identity was used to simplify the above.  We are now
in a situation very similar to that in Section~\ref{sec:dualLiegroup}. 
The difference is that the relevant metrics are now $\sigma_{ij}$ and
$\sigmatil_{ij}$.  The difficulty arises in that we have lost positive
definiteness of the metrics.  The only constraint is that $\sigma +
\sigmatil = I$.  If either $\sigma$ or $\sigmatil$ is definite then an
analysis along the lines of Section~\ref{sec:dualLiegroup} leads to
the conclusion that $\lieg$ and $\liegtil$ are abelian.  If both are
indefinite or both are singular then the analysis previously provided
breaks down. This situation also merits further study.

\section{Poisson-Lie duality}
\label{sec:Poisson-Lie}

Here we discuss a beautiful example of scenario~\oaref{item:symplectic}{3}
described at the end of Section~\oaref{sec:psi}{8.1} where we are given a
special symplectic bifibration and we have to construct the metrics and
antisymmetric tensors on $M$ and $\Mtil$.  The Drinfeld double Lie
group is an example of a special symplectic bifibration.  The metrics and
antisymmetric tensors constructed in this manner correspond to the
Poisson-Lie duality of Klimcik and Severa~\cite{Klimcik:1995ux}.  The
explicit duality transformation was obtained by
Sfetsos~\cite{Sfetsos:1998pi}.  We can use our general framework to
determine both by making educated guesses.  The Drinfeld double
$G_{D}$ is a Lie group whose Lie algebra $\lieg_{D}$ is a Lie
bialgebra, see Appendix~\ref{sec:bialgebra}.  The bifibration is by
Lie groups $G$ and $\Gtil$ with respective Lie algebras $\lieg$ and
$\tilde{\lieg} \approx \lieg^{*}$.  The Lie algebras are related by
$\lieg_{D} = \lieg \oplus \tilde{\lieg} = \lieg \oplus \lieg^{*}$.  If
$\{T_{a}\}$ is a basis for $\lieg$ and $\{\Ttil^{a}\}$ is the
associated dual basis for $\lieg^{*}$ then
\begin{eqnarray*}
    [T_{a},T_{b}] & = & C^{c}{}_{ab}T_{c}\,,  
    \\
    \relax [T_{a},\Ttil^{b}] & = & 
    \Ctil_{a}{}^{bc}T_{c}-C^{b}{}_{ac}\Ttil^{c}\, ,
    \\
    \relax [\Ttil^{a},\Ttil^{b}] & = & \Ctil_{c}{}^{ab}\Ttil^{c}\,,
\end{eqnarray*}
where $C^{c}{}_{ab}$ and $\Ctil_{a}{}^{bc}$ are respectively the
structure constants for $G$ and $\Gtil$.  The two sets of structure
constants must satisfy compatibility condition
(\ref{eq:compatability}).  To write down the symplectic structure in a
convenient way we introduce some notation slightly different than the 
one given in \cite{Klimcik:1995ux}.  Let
$g\in G$ then the adjoint representation on $\lieg$ is given by
$gT_{b}g^{-1}=T_{a}\,a^{a}{}_{b}(g)$.  One also has $g \Ttil^{a}g^{-1}
= a^{a}{}_{b}(g^{-1})(\Ttil^{b} + \Pi^{bc}(g)T_{c})$ where
$\Pi^{ab}=-\Pi^{ba}$.  Similarly one has that if $\gtil\in\Gtil$ then
$\gtil \Ttil^{b}\gtil^{-1} = \Ttil^{a}\, \atil_{a}{}^{b}(\gtil)$ and
$\gtil T_{a} \gtil^{-1} = \atil_{a}{}^{b}(\gtil^{-1})(T_{b} +
\Pitil_{bc}(\gtil) \Ttil^{c})$ where $\Pitil$ is antisymmetric.  It is
worthwhile to note that if $g=e^{x^{a}T_{a}}$ then $\Pi^{ab}(g) =
x^{c}\Ctil_{c}{}^{ab} + O(x^{2})$ and similarly for $\Pitil_{ab}$. 
Drinfeld shows that the
bivector $\Pi^{ab} T_{a}\wedge T_{b}$ on $G$ defines a Poisson bracket
that is compatible with the group multiplication law 
\cite{Drinfeld:83aa,Semenov:85aa}.  A Poisson-Lie
group is a Lie group with a Poisson structure which is compatible with
the group operation.  Thus we have that $G$ is a Poisson-Lie group. 
Note that the Poisson bivector is degenerate.  Similarly the bivector
$\Pitil_{ab}\Ttil^{a}\wedge\Ttil^{b}$ makes $\Gtil$ a Poisson Lie
group.  Klimcik and Severa discovered that the sigma model defined on
the Poisson-Lie group $G$ is dual to the sigma model defined on the
Poisson-Lie group $\Gtil$ hence the name Poisson-Lie duality.  To
exhibit the duality transformation we write down the symplectic
structure on $G_{D}$.  We note that the Drinfeld double is not a
Poisson-Lie group in the Poisson structure associated with the
symplectic structure since the Poisson bivector would be
nondegenerate.  In the (perfect) Drinfeld double every element $k\in
G_{D}$ can uniquely be written as $k= gu$ or $k=vh$ where $g,h\in G$
and $u,v\in\Gtil$.  The inverse function theorem shows that you can
choose $h$ and $u$ as local coordinates on $G_{D}$ near the identity. 
Let $\lambda = h^{-1}dh$ and $\lambdatil = u^{-1}du$ be respectively
the left invariant Maurer-Cartan forms on $G$ and $\Gtil$.  The
symplectic structure \cite{Alekseev:1994qs,Alekseevsky:98aa}
may be written as
\begin{eqnarray}
    2\beta & = &
    (du\,u^{-1})_{a}\wedge (g^{-1}dg)^{a} 
    + (v^{-1}dv)_{a}\wedge (dh\,h^{-1})^{a}\,,
    \nonumber \\
    & = & \lambdatil_{a}\wedge
    \left[ \lambda^{b} + \lambdatil_{c}\Pi^{cb}(h^{-1})\right]
    (I - \Pitil(u^{-1})\Pi(h^{-1}))^{-1}{}_{b}{}^{a}
    \nonumber \\
     & +  & \left[ \lambdatil_{b} + \lambda^{c}\Pitil_{cb}(u^{-1})\right]
     \wedge \lambda^{a}(I - 
     \Pi(h^{-1})\Pitil(u^{-1}))^{-1}{}^{b}{}_{a}\;.
     \label{eq:drinsymp}
\end{eqnarray}
Using $\beta$ we can construct the duality transformations. All we 
have to verify is that we get metrics and antisymmetric tensor fields 
on $G$ and $\Gtil$. 

Klimcik and Severa show that the symmetric and antisymmetric parts of
the rank two tensors $E = \lambda^{t}\otimes (F^{-1} +
\Pi)^{-1}\lambda$ and $\Etil = \lambdatil^{t}\otimes (F +
\Pitil)^{-1}\lambdatil$ are respectively the metrics and antisymmetric
tensors for the dual sigma models on $G$ and $\Gtil$.  The
coefficients $F_{ab}$ are constants.  By an appropriate choice of
basis for $\lieg$ one can always choose $F= I + b$ where $b$ is
antisymmetric.  For pedagogical reasons we first look at the special
case where $b=0$ which is analogous to choosing $n^{0}=0$ in
(\ref{eq:nintegral}).  By looking at (\ref{eq:drinsymp}) we make an
educated guess and conjecture that our orthonormal bases should be
given by $\theta$ in $G$ and $\thetatil$ in $\Gtil$ where
\begin{equation}
    \lambda = (I+\Pi)\theta\quad\mbox{and}\quad \lambdatil =
    (I+\Pitil)\thetatil\;.
    \label{eq:drinfeldortho}
\end{equation}
We can verify that this is in agreement with the Klimcik and Severa
data by noting that $E = \theta^{t}\otimes(I-\Pi)\theta$ and $\Etil=
\thetatil^{t}\otimes(I-\Pitil)\thetatil$.  The symmetric parts on $E$
and $\Etil$ in this basis are respectively $\theta^{t}\otimes\theta$
and $\thetatil^{t}\otimes\thetatil$ and thus we see that we have
orthonormal bases on $G$ and $\Gtil$.  Likewise we see that the
components of the antisymmetric tensors in this basis are given by
$-\Pi$ and $-\Pitil$ respectively.  Note that $\theta$ ($\thetatil$)
is well defined on $G$ ($\Gtil$) because $\Pi$ ($\Pitil$) is defined
on $G$ ($\Gtil$).

We are now ready to verify that there is a duality transformation. 
Postulate that frames given by $\theta$ and $\thetatil$ are the
orthonormal ones we need.  Rewrite (\ref{eq:drinsymp}) in the 
orthonormal frame where you find
\begin{eqnarray}
    m & = & (I-\Pitil)(I-\Pi\Pitil)^{-1}(I+\Pi)\,,
    \label{eq:drinm}  \\
    \ltil & = & -(I-\Pitil)\Pi(I-\Pitil\Pi)^{-1}(I+\Pitil)\,,
    \label{eq:drinltil}  \\
    l & = & -(I-\Pi)\Pitil(I-\Pi\Pitil)^{-1}(I+\Pi)\,,
    \label{eq:drinl}
\end{eqnarray}
using the notation in (\oaref{eq:betaframe}{3.5}). A brief computation shows 
that $m=I + n$ where $n$ is antisymmetric and given by
\begin{eqnarray*}
   n &=& \sum_{k=1}^{\infty}\left[(\Pi\Pitil)^{k}-(\Pitil\Pi)^{k}\right]
   \\
   &+& \sum_{k=0}^{\infty}(\Pi\Pitil)^{k}\Pi -
       \sum_{k=0}^{\infty}\Pitil(\Pi\Pitil)^{k}\;.
\end{eqnarray*}
In this frame $m$ is already in normal form and we can proceed.  Note
that $\ntil = -n$ as follows from (\oaref{eq:mn}{3.15}).  Using
(\oaref{eq:defn}{3.8}) and (\oaref{eq:defntil}{3.9}) 
we see that $B = l-m+I = -\Pi$
and $\Btil = \ltil + m -I = -\Pitil$.  The important result here is
that $B$ and $\Btil$ are quantities which respectively live on $G$ and
$\Gtil$ and thus we have constructed the Poisson-Lie duality of
Klimcik and Severa for the special case $b=0$. 

The general solution for arbitrary $b$ is given by choosing the 
orthonormal frames to be given by
\begin{equation}
    \lambda = (I+\Pi F)\theta\quad\mbox{and}\quad \lambdatil =
    (F+\Pitil)\thetatil\;.
    \label{eq:drinfeldorthogen}
\end{equation}
We have $E =\theta^{t}\otimes(F - F^{t}\Pi F)\theta$ and
$\Etil = \thetatil\otimes(F^{t}-\Pitil)\thetatil$ with the components
of the antisymmetric tensor fields given by $B=b-F^{t}\Pi F$ and
$\Btil = -b -\Pitil$ in this basis. The components of the symplectic 
form in this basis are
\begin{eqnarray}
    m & = & (F^{t}-\Pitil)(I-\Pi\Pitil)^{-1}(I+\Pi F)\,,
    \label{eq:drinmgen}  \\
    \ltil & = & -(F^{t}-\Pitil)\Pi(I-\Pitil\Pi)^{-1}(F+\Pitil)\,,
    \label{eq:drinltilgen}  \\
    l & = & -(I-F^{t}\Pi)\Pitil(I-\Pi\Pitil)^{-1}(I+\Pi F)\,,
    \label{eq:drinlgen}
\end{eqnarray}
A brief computation shows that $m=I+n$ where $n$ is antisymmetric and 
given by
\begin{eqnarray*}
   n &=& -b + \sum_{k=1}^{\infty}
       \left[F^{t}(\Pi\Pitil)^{k}-(\Pitil\Pi)^{k}F\right]
   \\
   &+& \sum_{k=0}^{\infty} F^{t}(\Pi\Pitil)^{k}\Pi F -
       \sum_{k=0}^{\infty}\Pitil(\Pi\Pitil)^{k}\;.
\end{eqnarray*}
Using (\oaref{eq:defn}{3.8}) and (\oaref{eq:defntil}{3.9}) 
we see that $B = l-m+I =
b- F^{t}\Pi F$ and $\Btil = \ltil + m -I = -b -\Pitil$.  We have
reproduced the ansatz of Klimcik and Severa and the duality 
transformation of Sfetsos.

\section{Infinitesimally homogeneous $n_{ij}$}

\subsection{General theory}
\label{sec:inf-hom-n}

In this section we address the question, ``What if $n_{ij}$ is the
same everywhere?''  We will see that this is a much weaker condition
than saying $n'=n''=0$ which we already studied in
Section~\oaref{sec:covconst}{8.2} and lead to abelian duality.  We will show
that $P$ is a homogeneous space under certain assumptions.  First we
have to address the question of what does ``same everywhere'' mean. 
The best way to do this is to exploit some ideas developed by 
Singer~\cite{Singer:60aa}
for the study of homogeneous spaces.  It is convenient to work in the
bundle $\framebundle$ of the adapted orthogonal frames we have been
using.  This bundle has structure group $\Orth(n)$ and it admits a
global coframing given by $(\theta^{i},\thetatil^{j},\psi_{kl})$ where
$(\theta^{i},\thetatil^{j})$ are the canonical $1$-forms on the frame
bundle and $\psi_{ij}$ is an $\Orth(n)$ connection.  Remember that the
Maurer-Cartan form on $\Orth(n)$ is the restriction of $\psi$ to a
fiber of $\framebundle$.  The relationships among the geometries of
$M,\Mtil,P$ are encapsulated in the Cartan structural equations for
$\framebundle$:
\begin{eqnarray}
    d\theta^{i} & = & -\psi_{ij}\wedge\theta^{j} - \half 
    f_{ijk}\theta^{j}\wedge\theta^{k}\;,
    \label{eq:cartan1psi1}  \\
    d\thetatil^{i} & = & -\psi_{ij}\wedge\thetatil^{j} - \half 
    \ftil_{ijk}\thetatil^{j}\wedge\thetatil^{k}\;,
    \label{eq:cartan1psitil1} \\
    d\psi_{ij} & = & -\psi_{ik}\wedge\psi_{kj} - 
    T''_{ijlm}\theta^{l}\wedge\thetatil^{m}\;.
    \label{eq:cartan2psi1}
\end{eqnarray}
where $f_{ijk}=-f_{ikj}$, $\ftil_{ijk}=-\ftil_{ikj}$ and
$T''_{ijkl}=-T''_{jikl}$.  Here $f,\ftil,T''$ are all functions on
$\framebundle$.  Note that there is torsion arising from the reduction
of the structure group.  The ideal generated by
$\{\theta^{1},\ldots,\theta^{n}\}$ is a differential ideal with
integral submanifolds being the restriction of $\framebundle$ to the
fibers of $\Pi$.  The ideal generated by
$\{\thetatil^{1},\ldots,\thetatil^{n}\}$ is a differential ideal with
integral submanifolds being the restriction of $\framebundle$ to the
fibers of $\Pitil$.  The degenerate quadratic forms
$\theta^{i}\otimes\theta^{i}$ and $\thetatil^{i}\otimes\thetatil^{i}$
on $\framebundle$ are respectively pullbacks of the metrics on $M$ and
$\Mtil$.  The pullback of the Riemannian connection on the frame
bundle of $M$ is schematically $\psi + f\theta$ and likewise for
$\Mtil$.  Said differently, when restricting $\psi_{ij}$ to a
``horizontal fiber'' you get an orthogonal connection on the fiber
with torsion, \emph{etc}.  We remind the reader that a tensor in $P$
is a collection of functions on $\framebundle$ which transform
linearly under the action of $\Orth(n)$ on $\framebundle$,
\emph{i.e.}, as you change frames the ``tensor'' transforms
appropriately.  For future use the frame dual to the coframe
$(\theta^{i},\thetatil^{j},\psi_{kl})$ will be denoted by
$(e_{i},\etil_{j},E_{kl})$.  The horizontal vector fields with respect
to $\psi$ are spanned by $\{e_{A}\}=\{e_{i},\etil_{j}\}$.

We are now ready to define the statement ``$n_{ij}$ is the same
everywhere''.  Pick a point $b\in\framebundle$.  If we go to a rotated
frame $Rb$, $R\in\Orth(n)$, then $n_{ij}(b)$ becomes
$n_{ij}(Rb)=R_{ik}R_{jl}n_{kl}(b)$.  Notice that as we move along the
fiber going through $b$ we will get the full orbit of $n_{ij}(b)$
under $\Orth(n)$.  Thus to make sense of ``$n_{ij}$ is the same
everywhere'' we should not really talk about $n_{ij}$ but about the
invariants of antisymmetric tensors under the orthogonal group.  We
should be thinking in terms of the space of orbits of antisymmetric
tensors under $\Orth(n)$.  In the frame bundle, the functions $n_{ij}$
define a map $n:\framebundle \to \bigwedge^{2}(\bbR^{n})$.  We say that
$P$ is $n$-\emph{homogeneous} if the image of the map $n:\framebundle
\to \bigwedge^{2}(\bbR^{n})$ is a single $\Orth(n)$-orbit.  Said
differently, you get the same $2 \times 2$ block diagonalization of
$n_{ij}$ at each point of $P$.  This is a weaker condition than
covariantly constant $n$.  If $P$ is simply connected then a
covariantly constant $n_{ij}$ is determined by parallel transporting
$n_{ij}$ from a reference point.  The value of $n$ at the reference
point determines $n$ everywhere.

The condition that $P$ be $n$-homogeneous is not strong enough for us. 
This leads to the notion of ``infinitesimally $n$-homogeneous'' where
not only is $n$ the same everywhere but also the first $(N+1)$
covariant derivatives of $n$.  Let $\nabla n$ denote the covariant
derivative of $n$, $\nabla^{2}n$ the second covariant derivative of
$n$, \emph{etc}.  Consider the map
$$
    \rho^{s}=(n,\nabla n, \nabla^{2}n,\ldots,\nabla^{s}n) : 
    \framebundle \to \bbR^{J_{s}}\;,
$$
where $J_{s}$ is an integer we do not compute.  We say that $P$ is
\emph{infinitesimally $n$-homogeneous} if image of the map
$\rho^{N+1}$ is a single $\Orth(n)$-orbit.  The integer $N$ is
determined inductively as follows.  

First we do a rough argument and afterwards we state Singer's result. 
Look at $\rho^{0}=n:\framebundle \to \bigwedge^{2}(\bbR^{n})$ and pick
a point $n^{0}$ in the orbit.  Consider $\calB =\{b\in\framebundle:
n(b)=n^{0}\}$.  Note that $\calB$ is a sub-bundle of $\framebundle$
because $n(\framebundle)$ is a single orbit.  If
$K'_{0}\subset\Orth(n)$ is the isotropy group of $n^{0}$ then the
action of $K'_{0}$ on a point $b\in\calB$ leaves you in $\calB$.  Thus
$\calB$ is a principal sub-bundle of $\framebundle$ with structure
group $K'_{0}$.  The choice of $n^{0}$ has broken the symmetry group
to $K'_{0}$.  Now let us be precise about Singer's result. 
Pick a
$b_{0}\in\framebundle$.  There exists a principal sub-bundle
$\calB_{0}\subset\framebundle$ containing $b_{0}$ such that $n$ is
constant on $\calB_{0}$ and the structure group $K_{0}\subset\Orth(n)$
of $\calB_{0}$ is the connected component of the isotropy group of
$n(b_{0})$.  Note that for a generic orbit, $K_{0}$ will be a maximal
torus of $\Orth(n)$.

Next we invoke $\nabla n$ to reduce the symmetry group some more.  We
use $\rho^{1}$ and apply Singer's theorem to it.  There exists a
principal sub-bundle $\calB_{1}\subset\calB_{0}$ containing $b_{0}$
such that $(n,\nabla n)$ is constant on $\calB_{1}$ and the structure
group $K_{1}\subset K_{0}$ of $\calB_{1}$ is the connected component
of the isotropy group of $(n(b_{0}),\nabla n(b_{0}))$.  We continue
the procedure by looking at $\rho^{2},\rho^{3},\ldots$ and finding a
sequence of principal sub-bundles $\framebundle \supset \calB_{0}
\supset \calB_{1} \supset \cdots \supset \calB_{N} \supset
\calB_{N+1}$ with respective structure groups $\Orth(n)\supset
K_{0}\supset K_{1} \supset \cdots \supset K_{N} \supset K_{N+1}$. 
Since $\Orth(n)$ is finite dimensional there exists an integer $N$
such that the chain of groups satisfies the property that $K_{0}\neq
K_{1}\neq \cdots \neq K_{N-1} \neq K_{N} = K_{N+1}$.  In fact Singer
establishes that $\calB_{N}=\calB_{N+1}$, henceforth denoted by $H$,
is a principal bundle with structure group $K=K_{N}$.  Our arguments
show that $\rho^{N+1}=(n,\nabla n,\ldots,\nabla^{N+1}n)$ is constant
on $H$.  Note that structure group $K$ is the connected component of
the isotropy group of $\rho^{N}(b_{0})$.  The chain of groups tells us
that $N \le \half n(n-1)$ and later on we will see that we also
require $N\ge 1$.

Next we show that $H$ is a Lie group and conclude that $P=H/K$ is a
homogeneous space.  The strategy is to write down the Cartan
structural equations for the principal bundle $H$ and show that they
are actually the Maurer-Cartan equations for a group.  Pick a point
$b\in\framebundle$ and observe that $d\rho^{N}$ is tangent to the
orbit because $\rho^{N}(\framebundle)$ is a single orbit.  The orbit
is generated by the action of $\Orth(n)$ therefore for $X\in
T_{b}\framebundle$ there exists a linear map $\Xi:T_{b}\framebundle
\to \so(n)$ such that $d\rho^{N}(X)=\Xi(X)\cdot \rho^{N}(b)$.  Use the
standard metric on $\SOrth(n)$ to write an orthogonal decomposition
$\so(n) = \liek \oplus \liek^{\perp}$ where $\liek$ is the Lie algebra
of $K$ and $\liek^{\perp}$ is its orthogonal complement.  Let
$b_{0}\in H$ then we observe that if $\Xi,\Xi'$ are such that for
$X\in T_{b_{0}}\framebundle$ you have
$d\rho^{N}(X)=\Xi(X)\cdot\rho^{N}(b_{0})=\Xi'(X)\cdot\rho^{N}(b_{0})$
then $\Xi'(X)-\Xi(X)\in\liek$.  At $b_{0}\in H$ you can uniquely
specify $\Xi$ by requiring that $\Xi(X)\in\liek^{\perp}$.  We will
make this choice.  Note that we allow $X$ to be in the full tangent
space $T_{b_{0}}\framebundle$.  In summary, for $b_{0}\in
H\subset\framebundle$ there exists a unique linear transformation
$\Xi:T_{b_{0}}\framebundle\to\liek^{\perp}$ such that
\begin{equation}
    d\rho^{N}(X)=\Xi(X)\cdot \rho^{N}(b_{0}).  
    \label{eq:Xi}
\end{equation}
The definition of the 
covariant derivative is
\begin{eqnarray}
    d\rho^{N}(X) &=& -\psi(X)\cdot\rho^{N}(b_{0}) 
    \nonumber \\
    &+& ((\nabla_{X}n)(b_{0}),\nabla_{X}(\nabla
    n)(b_{0}),\nabla_{X}(\nabla^{2}n)(b_{0}),\ldots,
    \nabla_{X}(\nabla^{N}n)(b_{0})).
    \label{eq:rho}
\end{eqnarray}
Under the decomposition $\so(n)=\liek\oplus\liek^{\perp}$ we have
$\psi = \psik + \psikperp$.  Upon restriction to $H$, $\psik$ is a
$K$-connection on the principal bundle $H$ and $\psikperp$ will become
torsion.  Since $K$ is the isotropy group of $\rho^{N}(b_{0})$ we
conclude that $\psik\cdot\rho^{N}(b_{0})=0$.  If we restrict
(\ref{eq:rho}) to $H\subset\framebundle$ and choose $X\in T_{b_{0}}H$
then $d\rho^{N}(X)=0$ because $\rho^{N}$ is constant on $H$.  Thus for
$X\in T_{b_{0}}H$ we have that
\begin{equation}
    \psikperp(X)\cdot\rho^{N}(b_{0}) = 
    ((\nabla_{X}n)(b_{0}),\nabla_{X}(\nabla
    n)(b_{0}),\nabla_{X}(\nabla^{2}n)(b_{0}),\ldots,
    \nabla_{X}(\nabla^{N}n)(b_{0})).
    \label{eq:psiperp}
\end{equation}
If we think of the above as a series of linear equations for 
$\psikperp(X)$ then it is easy to see that if a solution exists then 
in must be unique. Next we show that the solution exists. To do this
we observe that the covariant derivative (with 
connection $\psi$) of
$\rho^{N}$ in direction $e_{A}$ is given by $d\rho^{N}(e_{A}) =
\Xi(e_{A})\cdot\rho^{N}(b_{0})$, see (\ref{eq:Xi}), (\ref{eq:rho}). 
Thus we have
$$
\Xi(e_{A})\cdot\rho^{N}(b_{0}) = 
((\nabla_{X}n)(b_{0}),\nabla_{X}(\nabla
    n)(b_{0}),\nabla_{X}(\nabla^{2}n)(b_{0}),\ldots,
    \nabla_{X}(\nabla^{N}n)(b_{0})).
$$
Comparing this with (\ref{eq:psiperp}) and using the uniqueness of the 
solution we conclude that the torsion $\psikperp(e_{A})=\Xi(e_{A})$. 
Also note that the right hand side of (\ref{eq:psiperp}) is constant  
on $H$ and thus the $\psikperp(e_{A})$ must be constant on $H$ by 
uniqueness. On restriction to $H$ we have
\begin{equation}
    \psikperp_{ij} = \tor_{kij}\theta^{k} 
    +\tortil_{kij}\thetatil^{k}\;,
    \label{eq:ktorsion}
\end{equation}
where $\tor_{kij}$ and $\tortil_{kij}$ are constant on $H$.

We are now almost ready to write down the Cartan structural equations
for $H$.  First we observe that certain functions are constant.  If we
let $F$ denote $f_{ijk}$ or $\ftil_{ijk}$ then equations
(\oaref{eq:fnpsi}{8.17}) and (\oaref{eq:ftilnpsi}{8.18}) 
may schematically be written
$(I+n)F=\nabla n$.  By differentiating we learn that $\nabla^{s}F$ is
a function of $(n,\nabla n,\ldots,\nabla^{s+1}n)$ only.  If $N \ge 1$
then $f_{ijk}, \ftil_{ijk},T''_{ijkl}$ in (\ref{eq:cartan1psi1}),
(\ref{eq:cartan1psitil1}) and (\ref{eq:cartan2psi1}) are constant on
$H$ since $\rho^{N+1}$ is constant on $H$.  We require $N\ge 1$.  The
first Cartan structural equations (\ref{eq:cartan1psi1}) and
(\ref{eq:cartan1psitil1}) become
\begin{eqnarray}
    d\theta^{i} & = & -\psik_{ij}\wedge\theta^{j} - \half 
    c_{ijk}\theta^{j}\wedge\theta^{k} + 
    \tortil_{kij}\theta^{j}\wedge\thetatil^{k}\;,
    \label{eq:cartanH1}  \\
    d\thetatil^{i} & = & -\psik_{ij}\wedge\thetatil^{j} -\half 
    \ctil_{ijk}\thetatil^{j}\wedge\thetatil^{k} + 
    \tor_{kij}\thetatil^{j}\wedge\theta^{k}\;,
    \label{eq:cartantilH1}
\end{eqnarray}
where $c,\ctil,\tor,\tortil$ are constant and $c_{ijk}=-c_{ikj}$,
$\ctil_{ijk}=-\ctil_{ikj}$.  Think of the $ij$ indices of $T''_{ijlm}$
as taking values in the $\so(n)$ Lie algebra and denote the projection
of $-T''_{ijlm}$ onto $\liek$ by $K^{\liek}_{ijlm}$.  Note that
$K^{\liek}_{ijlm}$ is constant on $H$. The second Cartan structural
equation (\ref{eq:cartan2psi1}) may be written as
\begin{eqnarray}
    d\psik_{ij} & = & -\psik_{ik}\wedge\psik_{kj}
    -(\psik_{ik}\wedge\psikperp_{kj})^{\liek} 
    - (\psikperp_{ik}\wedge\psik_{kj})^{\liek}
    \nonumber  \\
     & - & (\psikperp_{ik}\wedge\psikperp_{kj})^{\liek} + 
     K^{\liek}_{ijlm}\theta^{l}\wedge\thetatil^{m}
    \label{eq:cartanH2},
\end{eqnarray}
where you substitute (\ref{eq:ktorsion}) for $\psikperp$ in the above. 
The important lesson is that $H$ has a coframing given by 
$(\theta,\thetatil,\psik)$ and that the structural equations 
(\ref{eq:cartanH1}), (\ref{eq:cartantilH1}) and (\ref{eq:cartanH2}) 
only involve constants and thus are the Maurer-Cartan equations for a 
Lie group. We have shown that if $P$ is infinitesimally 
$n$-homogeneous with $N\ge 1$ then $P$ is a homogeneous space $H/K$ 
where the Lie group $H$ is a sub-bundle of the frame bundle 
$\framebundle$.

\subsection{The case of $K=\{e\}$}
\label{sec:Kequale}

This is the situation where the residual symmetry group $K$ is broken
all the way down to the identity group $\{e\}$.  In this case $P=H$,
the symplectic manifold $P$ is a Lie group, and
$\liek^{\perp}=\so(n)$.  The Maurer-Cartan equations for $P$ are
\begin{eqnarray}
    d\theta^{i} & = & -\half c_{ijk}\theta^{j}\wedge\theta^{k} +
    \tautil_{kij}\theta^{j}\wedge\thetatil^{k}\;,
    \label{eq:MC1}  \\
    d\thetatil^{i} & = & -\half 
    \ctil_{ijk}\thetatil^{j}\wedge\thetatil^{k} + 
    \tau_{kij}\thetatil^{j}\wedge\theta^{k}\;,
    \label{eq:MC2}
\end{eqnarray}
where $c_{ijk}=-c_{ikj}$, $\ctil_{ijk}=-\ctil_{ikj}$.  Also
$\tau_{kij}$ and $\tautil_{kij}$ are antisymmetric under $i
\leftrightarrow j$ for arbitrary $i,j$ reflecting that
$\liek^{\perp}=\so(n)$.  Note that $\thetatil$ generates a
differential ideal and thus $\thetatil=0$ defines a fibration
$\Pitil:P \to \Mtil$ with fibers isomorphic to a Lie group $G$ with
structure constants $c_{ijk}$.  Likewise, $\theta$
generates a differential ideal and thus $\theta=0$ defines a
fibration $\Pi:P \to M$ with fibers isomorphic to a Lie group
$\Gtil$ with structure constants $\ctil_{ijk}$. We also remark that 
$G$ and $\Gtil$ are Lie subgroups of $P$. If $\tau=0$ then $G$ is 
a normal subgroup of $P$. If $\tautil=0$ then $\Gtil$ is a normal 
subgroup of $P$.

Let $(e_{i},\etil_{j})$ be the basis dual to
$(\theta^{i},\thetatil^{j})$.  The Maurer-Cartan equations may be
reformulated as the Lie algebra relations
\begin{eqnarray}
    [e_{i},e_{j}] & = & c_{kij}e_{k}\,
    \label{eq:ee}  \\ \relax
    [\etil_{i},\etil_{j}] & = & \ctil_{kij}\etil_{k}\,
    \label{eq:etiletil}  \\ \relax
    [e_{i},\etil_{j}] & = & \tau_{ikj}\etil_{k} -\tautil_{jki}e_{k}\;.
    \label{eq:eetil}
\end{eqnarray}
All the statements made in the previous paragraph also follow from 
the above.

Consider the left invariant vector fields $X=X^{i}e_{i}$,
$Y=Y^{i}e_{i}$.  Note that $\myLie{X}\thetatil^{i}
=-X^{k}\tau_{kij}\thetatil^{j}$.  Let $\tau_{ij}(X) = X^{k}\tau_{kij}$
then using the identity $[\myLie{X},\myLie{Y}]\thetatil^{i}
=\myLie{[X,Y]}\thetatil^{i}$ you obtain
$[\tau(X),\tau(Y)]=\tau([X,Y])$.  Thus we have a Lie algebra
representation $\tau:\lieg \to \so(n)$.  This means we have a
representation of $G$ by real orthogonal $n\times n$ matrices. 
Likewise, $\tautil:\liegtil\to\so(n)$ is a Lie algebra representation
and we have a representation of $\Gtil$ by real orthogonal matrices. 
This does not mean that $G$ is a compact group if $\tau\neq0$.  A
comment made in Appendix~\ref{sec:Liegroupprimer} tells us that
$\lieg/(\ker\tau)$ is a Lie algebra of the form ``compact semisimple
+ abelian''.  We know nothing at all about the ideal
$\ker\tau$ so we cannot make a stronger statement about the structure 
of $\lieg$.  Similar remarks apply to $\Gtil$.

A Drinfeld double $G_{D}$ admits the following geometric 
characterization:
\begin{enumerate}
    \item It is a Lie group of dimension $2n$ with a bi-invariant
    quadratic form of type $(n,n)$.
     \item It is a bifibration with the property that the fibers are
     isotropic submanifolds of $G_{D}$. The fibers are also isomorphic
     to Lie groups $G$ and $\Gtil$.
\end{enumerate}
If we apply the above to our situation by requiring that the quadratic
form $\thetatil^{i}\otimes\theta^{i} + \theta^{i}\otimes\thetatil^{i}$
be bi-invariant then we learn that $P=H$ is a Drinfeld double,
$\tau_{ijk}=c_{ijk}$, $\tautil_{ijk}=\ctil_{ijk}$,
$\tau_{kij}=-\tau_{ikj}$ and $\tautil_{kij}=-\tautil_{ikj}$.  It
follows that both $c_{ijk}$ and $\ctil_{ijk}$ are totally
antisymmetric and thus the quadratic forms
$\theta^{i}\otimes\theta^{i}$ and $\thetatil^{i}\otimes\thetatil^{i}$
give bi-invariant positive definite metrics on $G$ and $\Gtil$
respectively.  Thus $G$ and $\Gtil$ are of the type ``compact
semisimple + abelian''.  The symplectic structure on $P$ (if it
exists) appears to be different than the standard symplectic structure
on the Drinfeld double, see (\ref{eq:drinsymp}). In the examples I am 
familiar, if $G$ is simple and compact the its dual $\Gtil$ 
constructed via $R$-matrices is neither simple nor compact. I do not 
know what is known in this more general case ``compact semisimple + 
abelian''.

Returning to the general case we remark that the equations 
$d^{2}\theta=0$ and $d^{2}\thetatil=0$ leads to the conclusion that 
$c_{ijk}$ and $\ctil_{ijk}$ are respectively the structure constants 
for Lie groups $G$ and $\Gtil$, $\tau:\lieg\to\so(n)$ and 
$\tautil:\liegtil\to\so(n)$ are Lie algebra representations and
\begin{eqnarray}
    \tau_{kjm}\tautil_{jil} - \tau_{ljm}\tautil_{jik}
    +c_{ijk}\tautil_{mjl} - c_{ijl}\tautil_{mjk}
    + c_{jkl}\tautil_{mij}& = & 0\;,
    \label{eq:tautautil1}  \\
    \tautil_{kjm}\tau_{jil}-\tautil_{ljm}\tau_{jik}
    +\ctil_{ijk}\tau_{mjl} - \ctil_{ijl}\tau_{mjk}
    +\ctil_{jkl}\tau_{mij}& = & 0\;.
    \label{eq:tautatautil2}
\end{eqnarray}
These equations are generalizations of the corresponding equations
(\ref{eq:compatability}) in the bialgebra case.  These equations 
follow just from the structure equations for the group $P=H$.  There are
additional constraints which follow from duality considerations such
as $d\gamma = H-\Htil$ which lead to
\begin{eqnarray}
    c_{ijk}n_{il} + c_{ikl}n_{ij} + c_{ilj}n_{ik} & = & +H_{jkl}\,,
    \label{eq:Hcn}  \\
    \ctil_{ijk}n_{il} + \ctil_{ikl}n_{ij} + \ctil_{ilj}n_{ik} & = & 
    -\Htil_{jkl}\,,
    \label{eq:Htilctiln}  \\
    m_{jl}\tau_{kli}-m_{kl}\tau_{jli} 
    + n_{jl}\tautil_{ilk} -n_{kl}\tautil_{ilj} 
    & = & -m_{il}c_{ljk}\;,
    \label{eq:mtauc}  \\
    m_{jl}\tautil_{kli}-m_{kl}\tautil_{jli}
    +n_{jl}\tau_{ilk} - n_{kl}\tau_{ilj}
    & = & +\ctil_{ljk}m_{li}\;.
    \label{eq:mtautilctil}
\end{eqnarray}
We do not know if there are non-trivial solutions to these equations.

\section*{Acknowledgments}

I would like to thank O.~Babelon, L.~Baulieu, T.~Curtright,
L.A.~Ferreira, D.~Freed, S.~Kaliman, C-H~Liu, R.~Nepomechie,
N.~Reshetikhin, J.~S\'{a}nchez Guill\'{e}n, N.~Wallach and P.~Windey
for discussions on a variety of topics.  I would also like to thank
Jack Lee for his \emph{Mathematica} package Ricci that was used to
perform some of the computations.  I am particularly thankful to
R.~Bryant and I.M.~Singer for patiently answering my many questions
about differential geometry.  This work was supported in part by
National Science Foundation grant PHY--9870101.

\bigskip
\appendix
\par\noindent
{\bf\Large Appendices}
\section{Some Lie groups facts}
\label{sec:Liegroupprimer}

For clarification purposes we make some remarks about the left and
right actions on a Lie group.  For notational simplicity we restrict
to matrix Lie groups.  The identity element will be denoted by $I$. 
Let $G$ be a Lie group.  Let $a\in G$ then the left and right actions
on $G$ are respectively defined by $L_{a}g=ag$ and $R_{a}g = ga$ for
$g\in G$.  We take $\mathfrak{g}$, the Lie algebra of $G$, to be the
left invariant vector fields and we identify it with the tangent space
at the identity $T_{I}G$.  The left invariant Maurer-Cartan forms are
$\theta = g^{-1}dg$.  They satisfy the Maurer-Cartan equations
$d\theta = - \theta\wedge\theta$.  Pick a basis $(e_{1},\ldots,e_{n})$
of left invariant vector fields for $\mathfrak{g}$ with bracket
relations $[e_{i},e_{j}]=f^{k}{}_{ij}e_{k}$.  If the dual basis of
left invariant forms is $(\theta^{1},\ldots,\theta^{n})$ then
$d\theta^{i}=-\half f^{i}{}_{jk}\theta^{j}\wedge\theta^{k}$.

Naively you would expect that if $X$ is a left invariant vector field
then $\myLie{X}\theta =0$ since $\theta$ is left invariant.  In fact a
brief computation shows that $\myLie{X}\theta^{i} =
-(X^{k}f^{i}{}_{kj})\theta^{j}$ which is the adjoint action.  The
answer to this conundrum is that the left invariant vector fields
generate the right group action.  The easiest way to see this is to
use old fashioned differentials.  Let $v \in T_{I}G$ then the left
invariant vector field $X$ at $g$ which is $v$ at the identity is
given by $X=gv$.  The infinitesimal action of this vector field at $g$
is given by $g \to g + \epsilon X = g + \epsilon g v = g(I + \epsilon
v) \approx g e^{\epsilon v}$ which is the right action of the group. 
Thus we see that $g^{-1}dg \to e^{-\epsilon v}(g^{-1}dg)e^{\epsilon
v}$ which is the adjoint action in accordance with the Lie derivative
computation.  Take any metric $h_{ij}$ at the identity then
$h_{ij}\theta^{i}\otimes\theta^{j}$ is a left invariant metric on $G$. 
In general this metric is not invariant under the right action of the
group.  The right invariance condition is
$h_{il}f^{l}{}_{jk}+h_{jl}f^{l}{}_{ik}=0$ which means that the
structure constants are totally antisymmetric if the indices are
lowered using $h_{ij}$.  In such a situation the metric is
bi-invariant.

Assume you have a Lie algebra $\lieg$ with an invariant positive
definite metric then you have an orthogonal decomposition into ideals
$\lieg = \liek\oplus \mathfrak{a}$ where $\liek$ is a compact
semisimple Lie algebra and $\mathfrak{a}$ is abelian.  The proof is
straightforward and involves putting together a variety of
observations.  The invariance of the inner product $(\cdot,\cdot)$ is
the statement that the adjoint representation $\ad_{X}Y=[X,Y]$ is skew
adjoint with respect to the metric: $(\ad_{X}Y,Z)+(Y,\ad_{X}Z)=0$. 
The skew adjointness immediately leads to the decomposition of $\lieg$
into irreducible pieces.  It is an elementary exercise in linear
algebra to show that if $\mathfrak{h}$ is a non-trivial ideal in
$\lieg$ then its orthogonal complement $\mathfrak{h}^{\perp}$ is also
an ideal.  Since $\lieg=\mathfrak{h} \oplus \mathfrak{h}^{\perp}$ we
conclude that by continuing this process the Lie algebra decomposes
into irreducible ideals $\lieg=\lieg_{1} \oplus
\lieg_{2}\oplus\cdots\oplus\lieg_{N}$.  Let us collate all the abelian
subalgebras together into $\mathfrak{a}$ and rewrite the decomposition
as $\lieg = \liek_{1}\oplus\liek_{2} \oplus\cdots\oplus \liek_{M}
\oplus \mathfrak{a}$ where each $\liek_{j}$ is a simple lie algebra. 
Since this is a decomposition into ideals we have that $\mathfrak{a}$
is the center of the Lie algebra.  Let $(\cdot,\cdot)_{j}$ be the
restriction of the invariant inner product to $\liek_{j}$.  An
application of Schur's Lemma tells us that an invariant bilinear form
on a simple Lie algebra is a multiple of the Killing form.  Thus we
conclude that $(\cdot,\cdot)_{j} = \lambda K_{j}(\cdot,\cdot)$ where
$\lambda$ is a non-zero scalar and $K_{j}$ is the Killing
form\footnote{The sign of the Killing form is chosen such that it is
positive on a compact simple Lie algebra.} on $\liek_{j}$.  Since
$(\cdot,\cdot)$ is positive definite, the Killing form $K_{j}$ must be
definite and this is only possible if the Lie algebra $\liek_{j}$ is
of compact type.  This concludes the proof of the proposition in the
opening sentence.

Closely related to the above is the following.  If a Lie algebra
$\lieg$ has a faithful representation $\tau:\lieg\to\so(n)$ by skew
adjoint matrices then $\lieg = \liek \oplus \mathfrak{a}$ where
$\liek$ is a compact semisimple Lie algebra and $\mathfrak{a}$ is
abelian.  The proof follows from the observation that because the
representation is faithful we can think of $\lieg$ as a matrix Lie
subalgebra of $\so(n)$.  We know that $\so(n)$ has a positive definite
invariant metric so restriction to $\lieg$ induces a positive definite
invariant metric on $\lieg$.  We now use the proposition from the
previous paragraph.

We remark that if the representation $\tau$ in the 
previous paragraph is not faithful then $(\ker\tau) \subset 
\lieg$ is a nontrivial ideal in $\lieg$. The Lie algebra 
$\lieg/(\ker\tau)$ is of the form $\liek\oplus\mathfrak{a}$ but we can 
say nothing about the Lie algebra $(\ker\tau)$.

It is easy to see that the space of all invariant positive definite
metrics on a Lie algebra is a convex set. 
In a simple Lie algebra $\lieg$, the third cohomology
group $H^{3}(\lieg)$ is one dimensional and generated by the three
form $\omega(X,Y,Z) = K(X,[Y,Z])$ where $K$ is the killing form which
may be written in terms of the structure constants as
$\omega_{ijk}=K_{il}f^{l}{}_{jk}$.  If $h$ is an invariant metric then
$h(X,[Y,Z])$ is also a closed three form and by the convexity of the
space of positive definite invariant metrics it must be in the same
cohomology class as $\omega$.

\section{A primer on classical $R$-matrices}
\label{sec:primer}

There are two main nonequivalent approaches to classical $R$-matrices. 
The more familiar one is due to Drinfeld and based on the study of Lie
bialgebras~\cite{Drinfeld:83aa}.  The other due to
Semenov-Tian-Shansky is based on double Lie
algebras~\cite{Semenov:83aa} is the one directly related to our work. 
Here we discuss the interconnections between these two approaches. 
For an introduction to $R$-matrices and Poisson-Lie groups see the
book by Chari and Pressley~\cite{Chari:94aa} or the
article~\cite{Alekseevsky:98aa}.

\subsection{The Drinfeld Approach}
\label{sec:bialgebra}

Drinfeld bases his approach on the notion of a Lie bialgebra.  We 
begin with a down to earth approach.  Assume you have a Lie algebra 
$\lieg$ with basis $(e_{1},\ldots,e_{n})$ and Lie bracket relations 
$[e_{a},e_{b}] = f^{c}{}_{ab}e_{c}$.  If $X,Y \in \lieg$ then the 
adjoint action by $X$ is given by $\ad_{X}Y=[X,Y]$.  The adjoint 
action extends naturally to tensor products of $\lieg$.  Let 
$\lieg^{*}$ be the vector space dual with corresponding basis 
$(\omega^{1},\ldots,\omega^{n})$.  The Lie algebra $\lieg$ acts on 
$\lieg^{*}$ via the coadjoint representation $\ad^{*}_{e_{a}} 
\omega^{b} = -f^{b}{}_{ac}\omega^{c}$.  In general $\lieg^{*}$ is not 
a Lie algebra but there is a natural Lie algebra structure on 
$\lieg\oplus\lieg^{*}$ given by
\begin{eqnarray*}
    [e_{a},e_{b}] & = & f^{c}{}_{ab}e_{c}\,,  \\
    \relax [e_{a},\omega^{b}] & = & -f^{b}{}_{ac}\omega^{c}\,  \\
    \relax [\omega^{a},\omega^{b}] & = & 0\,.
\end{eqnarray*}
The most famous example is combining $\lieg=\so(3)$ and its Lie 
algebra dual $\lieg^{*}\approx\bbR^{3}$ into the Lie algebra of the 
euclidean group $E(3)$.  The situation becomes much more interesting 
when $\lieg^{*}$ is a Lie algebra in its own rights 
$[\omega^{a},\omega^{b}] = \fhat_{c}{}^{ab}\omega^{c}$.  You observe 
that $\lieg^{*}$ acts via its coadjoint action on its dual 
$(\lieg^{*})^{*}\approx\lieg$.  Thus one can consider the following 
more symmetric structure which takes into account the respective coadjoint 
actions
\begin{eqnarray}
    [e_{a},e_{b}] & = & f^{c}{}_{ab}e_{c}\,,  
    \label{eq:bialg1} \\
    \relax [e_{a},\omega^{b}] & = & 
    \fhat_{a}{}^{bc}e_{c}-f^{b}{}_{ac}\omega^{c}\, ,
    \label{eq:bialg2} \\
    \relax [\omega^{a},\omega^{b}] & = & \fhat_{c}{}^{ab}\omega^{c}\,.
    \label{eq:bialg3} 
\end{eqnarray}
This will be a  Lie algebra if the 
following conditions are satisfied:
\begin{equation}
    f^{e}{}_{ab}\fhat_{e}{}^{cd} =
    \fhat_{b}{}^{ed}f^{c}{}_{ae} + \fhat_{b}{}^{ce}f^{d}{}_{ae}
    -\fhat_{a}{}^{ed}f^{c}{}_{be}-\fhat_{a}{}^{ce}f^{d}{}_{be}\;.
    \label{eq:compatability}
\end{equation}
According to Drinfeld a \emph{Lie bialgebra} is a Lie algebra 
with Lie brackets (\ref{eq:bialg1}), (\ref{eq:bialg2}) and 
(\ref{eq:bialg3}).

Drinfeld gives a more abstract formulation which is more suitable for 
studying the abstract properties of a bialgebra and seeing the origins 
of classical $R$-matrices.  Assume you have a Lie algebra $\lieg$ and 
a ``cobracket'' $\Delta: \lieg \to \bigwedge^{2}\lieg$.  Drinfeld 
requires that the cobracket defines a Lie algebra on $\lieg^{*}$.  The 
structure constants on $\lieg^{*}$ are related to the cobracket by 
$\Delta(e_{a}) = \half \fhat_{a}{}^{bc}e_{b}\wedge e_{c}$.  
Compatibility condition (\ref{eq:compatability}) is incorporated via a 
cohomological argument.  The complex in question is
$$
\bigwedge\nolimits^{2}\lieg \longrightarrow \lieg^{*}\otimes 
\bigwedge\nolimits^{2}\lieg \stackrel{\coboundary}{\longrightarrow}
\bigwedge\nolimits^{2}\lieg^{*} \otimes \bigwedge\nolimits^{2}\lieg 
\longrightarrow \cdots\;.
$$
The coboundary operator (differential) is given by
\begin{equation}
    (\coboundary\Delta)(X,Y) = 
    \ad_{X}(\Delta(Y)) - \ad_{Y}(\Delta(X)) - \Delta([X,Y])  \,.
    \label{eq:coboundary}
\end{equation}
The condition (\ref{eq:compatability}) that glues the Lie algebras
into a Lie bialgebra is seen to be equivalent to the cocycle condition
$\coboundary\Delta=0$.  In this language, a \emph{Lie bialgebra} is a
Lie algebra $\lieg$ along with a cobracket $\Delta$ such that
$\lieg^{*}$ a Lie algebra and the cobracket is a $1$-cocycle.

A Lie bialgebra is exact if the cocycle is exact.  This means that 
there exists a $r\in \bigwedge^{2}\lieg$ such that $\Delta(X) = \coboundary 
r= \ad_{X}r$.  A computation shows that $r$ defines a bialgebra 
structure if and only if the Schouten bracket $\sleft r,r\sright$ is 
$\ad(\lieg)$-invariant: $\sleft r,r\sright \in 
(\bigwedge^{3}\lieg)_{\rm inv}$.  The Schouten bracket is defined by
$$
\sleft W\wedge X, Y\wedge Z \sright = [W,Y]\wedge X\wedge Z -
[W,Z]\wedge X\wedge Y -[X,Y]\wedge W\wedge Z + [X,Z]\wedge W\wedge Y.
$$
The condition $\sleft r,r \sright \in (\bigwedge^{3}\lieg)_{\rm inv}$ 
is called the modified classical Yang-Baxter equation (MCYBE) and 
$\sleft r,r\sright =0$ is called the classical Yang-Baxter equation 
(CYBE).

If $\lieg$ is semisimple then the Whitehead lemma states that
$H^{1}(\lieg,\bigwedge^{2}\lieg)=0$ and thus the cocycle $\Delta$ is
always a coboundary $\Delta = \coboundary r$.  Thus in this case we need to
understand the set of all $r\in \bigwedge^{2}\lieg$ which satisfy
MCYBE.

If $\lieg$ is simple then $(\bigwedge^{3}\lieg)_{\rm inv}$ is one
dimensional and is generated by the three index tensor obtained by
raising two indices on the structure constants using the Killing
metric.  If we call this object $B_{\rm K}$ then MCYBE becomes $\sleft
r,r\sright =aB_{\rm K}$ for some constant $a$.

Let us work in a basis.  If $r = \half r^{ab} e_{a}\wedge e_{b}$.  
Then $\Delta(e_{a}) = \ad_{e_{a}}r = \half r^{bc} 
\ad_{e_{a}}(e_{b}\wedge e_{c}) = \half \left( 
r^{dc}f^{b}{}_{ad}+r^{bd}f^{c}{}_{ad}\right) e_{b}$.  This tells us 
that
\begin{equation}
    \fhat_{a}{}^{bc}= r^{dc}f^{b}{}_{ad}+r^{bd}f^{c}{}_{ad}\;.
\end{equation}
Note that $\fhat_{a}{}^{bc}=-\fhat_{a}{}^{cb}$ because $r$ is skew.

It is possible to generalize the above by allowing the cocycle (now 
called $C$) to be in $\lieg\otimes\lieg$.  If you write $C = 
C^{ab}e_{a}\otimes e_{b}$ and you let $\Delta(X) = \ad_{X}C$.  In 
this case you find you get a Lie bialgebra if the following two 
conditions are satisfied:
\begin{enumerate}
    \item  $(C^{ab}+C^{ba})e_{a}\otimes e_{b}$, the symmetric part 
    of $C$, is $\ad(\lieg)$-invariant,

    \item   $\sleft C,C\sright \equiv [C^{12},C^{13}] = 
    [C^{12},C^{23}] + [C^{13},C^{23}] \in 
    \lieg\otimes\lieg\otimes\lieg$ is $\ad(\lieg)$-invariant.
\end{enumerate}
We use standard quantum group notation where for example 
$C^{13}=C^{ab}e_{a}\otimes I \otimes e_{b}$, \emph{etc}.  The bracket 
$\sleft\cdot,\cdot\sright$ above reduces to the Schouten bracket if 
$C$ is skew symmetric.  
We remark that the equation $\sleft C,C\sright=0$ is 
also called the classical Yang-Baxter equation. $C$ or $r$ are called 
classical $R$-matrices (by Drinfeld). The modified classical 
Yang-Baxter equation is $\sleft C,C\sright \in 
(\lieg\otimes\lieg\otimes\lieg)_{\rm inv}$.

A brief computation shows that
\begin{equation}
    \fhat_{a}{}^{bc}= C^{dc}f^{b}{}_{ad}+C^{bd}f^{c}{}_{ad}\;.
    \label{eq:ftoC}
\end{equation}
Let us write $C = \half s^{ab} (e_{a}\otimes e_{b} + e_{b}\otimes
e_{a}) + \half r^{ab}e_{a}\wedge e_{b}$ where $s$ is symmetric and $r$
is antisymmetric.  Since $s^{ab}$ is an $\ad(\lieg)$-invariant tensor
we have that 
\begin{equation}
    \fhat_{a}{}^{bc} = r^{dc}f^{b}{}_{ad}+r^{bd}f^{c}{}_{ad}\;.
    \label{eq:ftor}
\end{equation}
Thus the Lie algebra structure on $\lieg^{*}$ only depends on the 
antisymmetric part of $C$.  Note that $\fhat_{a}{}^{bc}$ will be skew 
under $b \leftrightarrow c$ as required.  The effect of the symmetric 
part $s_{ab}$ is to change the $\ad{\lieg}$-invariant term in the 
right hand side of the MCYBE. Equation (\ref{eq:ftoC}) may 
be interpreted as giving a Lie algebra homomorphism $C:\lieg^{*}\to\lieg$.

\subsection{Semenov-Tian-Shansky approach}
\label{sec:STS}

Semenov-Tian-Shansky's approach is directly influenced by classical 
integrable models where he needs that a single Lie algebra admits two 
different Lie brackets.  Let $\lieg$ be a Lie algebra and let 
$R:\lieg\to\lieg$ be a linear transformation (not necessarily 
invertible).  Define a skew operation $[\cdot,\cdot]_{R}$ by
\begin{equation}
     [X,Y]_{R} = [RX,Y] + [X,RY]\,.
    \label{eq:Rbracket}
\end{equation}
If $[\cdot,\cdot]_{R}$ is a Lie bracket then $R$ is called a classical 
R-matrix by Semenov-Tian-Shansky.  The Jacobi identity for 
$[\cdot,\cdot]_{R}$ may be written as
$$
    \bigg[X,[RY,RZ] - R([Y,Z]_{R})\bigg] + \textrm{cyclic permutations}=0
$$
A Lie algebra $\lieg$ with two Lie brackets $[\cdot,\cdot]$ and 
$[\cdot,\cdot]_{R}$ is called a \emph{double Lie algebra} by 
Semenov-Tian-Shansky.  The equation is $[RY,RZ] - R([Y,Z]_{R})=0$ is 
also called the CYBE. The equation
\begin{equation}
    [RY,RZ] - R([Y,Z]_{R})=-c[Y,Z]
    \label{eq:STSMYB}
\end{equation}
where $c$ is a constant is also called the MCYBE. Solutions to either
of these satisfy the Jacobi identity displayed above.

In a basis we have that $Re_{a}=e_{b}R^{b}{}_{a}$, 
$[e_{a},e_{b}]_{R}=(f^{c}{}_{ab})_{R}e_{c}$ and consequently 
(\ref{eq:Rbracket}) becomes $ [e_{a},e_{b}]_{R}= \left( R^{d}{}_{a} 
f^{c}{}_{db} +R^{d}{}_{b}f^{c}{}_{ad}\right) e_{c} $.  Thus the new 
structure constants are
\begin{equation}
    (f^{c}{}_{ab})_{R} = R^{d}{}_{a} 
    f^{c}{}_{db} +R^{d}{}_{b}f^{c}{}_{ad}.
    \label{eq:deffR}
\end{equation}
The Semenov-Tian-Shansky version of the MCYBE (\ref{eq:STSMYB}) is 
\begin{equation}
     f^{c}{}_{de}R^{d}{}_{a}R^{e}{}_{b}
    -f^{e}{}_{db}R^{d}{}_{a}R^{c}{}_{e} 
    -f^{e}{}_{ad}R^{d}{}_{b}R^{c}{}_{e}
    = -c f^{c}{}_{ab}\;.
    \label{eq:STSYBcomp}
\end{equation}

\subsection{Relating Drinfeld and Semenov-Tian-Shansky}
\label{sec:STStoD}

To relate the Semenov-Tian-Shansky approach and the Drinfeld approach
one needs an $\ad(\lieg)$-invariant metric on $\lieg$.  If $\lieg$ is
semisimple then one can take the $\ad(\lieg)$-invariant metric to be
the Killing metric.  The $\ad(\lieg)$-invariant metric is used to
identify $\lieg$ with $\lieg^{*}$.  By lowering indices $f_{abc}$ is
completely antisymmetric (due to the $\ad(\lieg)$-invariance).  We
wish to identify $f_{R}$ with $\ftil$.  Note that by rearranging
indices we have
\begin{eqnarray}
    (f_{c}{}^{ab})_{R} &= &R^{da}f_{cd}{}^{b} + R^{db}f_{c}{}^{a}{}_{d}
    \nonumber \\
    & = & R^{da}f^{b}{}_{cd} - R^{db}f^{a}{}_{cd}
    \label{eq:Rf}
\end{eqnarray}
If we are in a situation where $R^{ab}= - r^{ab}$ then we have related
$f_{R}$ to $\fhat$.  Said differently $R:\lieg\to\lieg$ is a
skew-adjoint operator with respect to the invariant metric.  In fact
there is a theorem \cite{Semenov:83aa} which states that if $\lieg$
has an $\ad(\lieg)$-invariant metric and if $R:\lieg\to\lieg$ is
skew-adjoint then the double Lie algebra is isomorphic to a Lie
bialgebra and all the structures in the Semenov-Tian-Shansky approach
(CYBE, MYBE) go into the structures in the Drinfeld approach (CYBE,
MYBE).  The isomorphism is given by thinking of the metric as giving a
map $\lieg\to\lieg^{*}$, \emph{i.e.}, lowering/raising indices.  We
are in a different situation because not all our $R$-matrices are
skew adjoint.

\providecommand{\href}[2]{#2}\begingroup\raggedright\endgroup

\end{document}

%% file: Ricci.tex
\def\mysavedown#1{\edef\mysubs{\mysubs#1}}
\def\mysaveup#1{\edef\mysups{\mysups#1}}
\def\mydown#1{{\mytensor}_{\vphantom{\mysubs}#1}}
\def\myup#1{{\mytensor}^{\vphantom{\mysups}#1}}
\def\tensor#1#2{
  #1
  \def\mytensor{\vphantom{#1}}
  \def\mysubs{\relax}
  \def\mysups{\relax}
  \let\down=\mysavedown
  \let\up=\mysaveup
  #2
  \let\down=\mydown
  \let\up=\myup
  #2
  }
